\documentclass[review]{elsarticle}

\usepackage{hyperref}
\usepackage{mathtools}
\usepackage{amsmath}
\newcommand\norm[1]{\left\lVert#1\right\rVert}
\DeclareMathOperator*{\argmin}{argmin}
\usepackage{algorithm}
\usepackage{algorithmic}
\usepackage{setspace}
\usepackage[utf8]{inputenc}
\usepackage{fourier} 
\usepackage{array}
\usepackage{makecell}
\usepackage{subcaption}
\usepackage{soul}

\restylefloat{algorithm}

\journal{Journal of \LaTeX\ Templates}

\begin{document}

\begin{frontmatter}

\title{PLIT: An alignment-free computational tool for identification of long non-coding RNAs in plant transcriptomic datasets}


\author[mymainaddress]{Sumukh Deshpande\corref{mycorrespondingauthor}}
\cortext[mycorrespondingauthor]{Corresponding author}
\ead{deshpan4@uni.coventry.ac.uk}

\author[mymainaddress]{James Shuttleworth}

\author[mymainaddress]{Jianhua Yang}

\author[mymainaddress]{Sandy Taramonli}

\author[mymainaddress]{Matthew England}

\address[mymainaddress]{School of Computing, Electronics and Mathematics, 1 Gulson Road Coventry University, Coventry, Warwickshire, CV1 2JH, United Kingdom}

\begin{abstract}
Long non-coding RNAs (lncRNAs) are a class of non-coding RNAs which play a significant role in several biological processes. RNA-seq based transcriptome sequencing has been extensively used for identification of lncRNAs. However, accurate identification of lncRNAs in RNA-seq datasets is crucial for exploring their characteristic functions in the genome as most coding potential computation (CPC) tools fail to accurately identify them in transcriptomic data. Well-known CPC tools such as CPC2, lncScore, CPAT are primarily designed for prediction of lncRNAs based on the GENCODE, NONCODE and CANTATAdb databases. The prediction accuracy of these tools often drops when tested on transcriptomic datasets. This leads to higher false positive results and inaccuracy in the function annotation process. In this study, we present a novel tool, PLIT, for the identification of lncRNAs in plants RNA-seq datasets. PLIT implements a feature selection method based on \textit{\(L_{1}\)} regularization and iterative Random Forests (iRF) classification for selection of optimal features. Based on sequence and codon-bias features, it classifies the RNA-seq derived FASTA sequences into coding or long non-coding transcripts. Using \textit{\(L_{1}\)} regularization, 31 optimal features were obtained based on lncRNA and protein-coding transcripts from 8 plant species. The performance of the tool was evaluated on 7 plant RNA-seq datasets using 10-fold cross-validation. The analysis exhibited superior accuracy when evaluated against currently available state-of-the-art CPC tools. 
\end{abstract}


\end{frontmatter}


\section{Introduction}

Recent advances in genome sequencing have led to the discovery of thousands of non-coding RNA transcripts. Using RNA Sequencing (RNA-Seq) and epigenome sequencing, a new class of RNA transcripts i.e. long non-coding RNAs (lncRNAs) is defined as those having transcript length $>$ 200 nucleotides. Although this class of RNA lacks protein-coding ability, they have been found involved in the regulation of biological processes such as enzymatic activity regulation, genomic loci imprinting, transcription, translation, and cellular differentiation \cite{Liu15}. Several lncRNA databases such as GENCODE, NONCODE and CANTATAdb have been developed for storage of lncRNAs \cite{Harrow12,YZhao16,Szczesniak2016}. These databases provide valuable resources for further identification of novel lncRNAs from genomic sequences. Although many lncRNAs have been identified in plants and animals, accurate computational identification of these lncRNAs in RNA-seq datasets remains one of the major problems in plants.

Therefore, an efficient, accurate and robust computational algorithm is required to predict lncRNAs in plants to further investigate their potential roles. Computational prediction of lncRNAs has been viable for the past few years. These methods generally use machine learning approaches to classify RNAs into different classes. Several tools have been developed including: Coding Potential Calculator 2 (CPC2) \cite{Kang2017}, Coding-Non-Coding Index (CNCI) \cite{Sun13}, Coding Potential Assessment Tool (CPAT) \cite{Wang13}, and a predictor of long non-coding RNAs and messenger RNAs based on improved k-mer scheme (PLEK) \cite{Li14}. CPC2 computes the coding probability of the sequence by computing its peptide length, isoelectric point, Fickett score \cite{Fickett82} and ORF integrity. CPC2 employed SVM using RBF kernel for training 17984 protein-coding and 10452 non-coding transcripts from Refseq \cite{OLeary16}, Ensembl (v87), and EnsemblPlants (v32) databases \cite{Zerbino2017}. Tools  such as CPAT and lncScore \cite{Wang13,JZhao16} classified protein-coding and non-coding transcripts based on logistic regression model as machine learning classifier using sequence-based features such as open-reading frame (ORF) size, ORF length, ORF coverage, GC content, Fickett score and hexamer score; whereas others such as CNCI and LncRNA-MFDL \cite{Fan15} classified lncRNAs using adjoining nucleotide triplets (ANT) features to identify most-like CDS (MLCDS) regions in each transcript. PLEK \cite{Li14} uses calibrated k-mer frequencies of a sequence and sliding-window approach as features for classification using SVM classifier from LIBSVM package. Currently developed alignment-free tools such as CPC2, lncScore, CPAT and PLEK work well with FASTA sequences derived from the GENCODE, NONCODE or CANTATAdb databases, but perform poorly on FASTA sequences derived from RNA-seq data. Thus, an accurate tool is required for prediction of lncRNAs in plants. 

In this work, we have developed a new alignment-free tool named \textbf{P}lant \textbf{L}ncRNA and \textbf{I}dentification \textbf{T}ool (PLIT) which uses \textit{\(L_{1}\)} regularization for feature selection and a Random Forest classifier for classification of sequences. For lncRNA identification, PLIT implements 73 sequence and codon-bias based features. The framework implements an optimization module called LASSO iterative Random Forest-Feature Selection (LiRFFS) \cite{Tibshirani96,Basu2018} which selects an optimal feature set from training and validation set features. The selected feature set can be used for identification of lncRNAs directly from RNA-seq derived FASTA sequences. The optimal features were selected based on coding and long non-coding FASTA sequences from the Refseq database \cite{OLeary16}. The prediction accuracy of the PLIT was benchmarked against other existing tools. In total, 31 features which included ORF length, ORF coverage, Hexamer Score, GC content, and codon-bias features such as Codon Usage Bias, Relative Codon Bias, and Relative Synonymous Codon Usage were selected. Compared with the existing tools, PLIT exhibited 15-30\% increase in the prediction accuracy when evaluated with 10-fold Cross Validation and repeated 10-fold Cross Validation with data shuffling on different plant RNA-seq datasets. The availability of the RNA-seq based plant lncRNA prediction tool will provide a useful resource for identification of novel lncRNAs in plants. PLIT is freely available on GitHub: https://github.com/deshpan4/PLIT. 

\section{Methods}

\subsection{Data description}

For extracting optimal feature set from FASTA sequences, a random selection of protein-coding and lncRNA transcript sequences from eight plant species was obtained from Refseq Release 91 \cite{OLeary16}. Transcript sequences for \textit{Arabidopsis thaliana}, \textit{Brassica rapa}, \textit{Brassica napus}, \textit{Brassica oleracea}, \textit{Zea mays}, \textit{Oryza sativa}, \textit{Solanum tuberosum} and \textit{Solanum lycopersicum} were downloaded. lncRNA sequences were filtered by applying a threshold cutoff of 200bp on non-coding RNA (ncRNA) FASTA files.

For performance evaluation, the lncRNA genomic coordinates for seven plant species (\textit{Arabidopsis Thaliana}, \textit{Glycine Max}, \textit{Oryza Sativa}, \textit{Solanum Lycopersicum}, \textit{Sorghum Bicolor}, \textit{Vitis Vinifera}, and \textit{Zea Mays}) were downloaded from CANTATAdb v2.0 \cite{Szczesniak2016} as negative examples, and protein-coding coordinates were downloaded from Ensembl Plants Release 41 \cite{Zerbino2017} as positive examples. The RNA-seq data for the seven plant species were obtained from the NCBI SRA database \cite{Leinonen2011} with accession numbers PRJNA268115, PRJNA237837, PRJNA293380, PRJNA478448, PRJNA318972, PRJNA356948 and PRJNA484195. A description of the total number of lncRNA transcript sequences and number of annotated sequences has been provided in Table 1.

\begin{table}[h!]
\vspace{-0.2cm}
\centering
\caption{Summary of lncRNAs used in plant RNA-Seq datasets.}
\begin{tabular}{l@{\hspace{0.3cm}}l@{\hspace{0.6cm}}l@{\hspace{0.4cm}}l@{\hspace{0.4cm}}l}
  \hline
Data set & lncRNAs used & lncRNAs annotated & Read length & Coverage\\
  \hline
\textit{A. thaliana} & 4373 & 1027 & 50-51 & 23x\\
\textit{G. max} & 2819 & 612 & 178-202 & 4x\\
\textit{O. sativa} & 2759 & 265 & 250 & 12x\\ 
\textit{S. lycopersicum} & 4308 & 549 & 250 & 6x\\
\textit{S. bicolor} & 2456 & 189 & 202 & 4x\\
\textit{V. vinifera} & 4019 & 1292 & 105 & 7x\\
\textit{Z. mays} & 10761 & 1029 & 202 & 1x\\
 \hline
\end{tabular}
\label{Tab:01}
\end{table}

\subsection{Data preprocessing}

The first 15 base pairs of the sequence reads are trimmed using Cutadapt \cite{Martin2011} to remove adapter and low-quality sequences with Q-score less than or equal to 30. For \textit{A. thaliana} reads, trimmed reads are aligned to the TAIR10 genome using the Tophat2 mapper \cite{Kim2013} with custom parameter values (minimum intron length = 40, maximum intron length = 5000, segment length = 20, segment mismatches = 1, max multihits = 1). The trimmed sequence reads of \textit{G. Max} were mapped to Glycine\_Max\_v2.0 genome with custom parameter values (minimum intron length = 30, maximum intron length = 15000, segment mismatches = 1, max multihits = 1). Trimmed reads for \textit{Oryza Sativa} L. ssp. Japonica were mapped to IRGSP-1.0 genome with min intron length of 20, max intron length of 15000, segment mismatches of 1 and max multihits of 1. For \textit{S. Lycopersicum}, the trimmed reads were aligned to the tomato genome (SL2.50) with min anchor length more than 8 nt, segment mismatches = 1 and max multihits = 1. The trimmed reads for \textit{Sorghum Bicolor} were mapped to the Sorgum genome (Sorghum\_Bicolor\_NCBIv3) and trimmed reads of \textit{Vitis Vinifera} were aligned to grape genome (IGGP\_12x) with segment mismatches = 1 and max multihits = 1. For \textit{Zea Mays}, the trimmed reads were aligned to the Maize genome (AGPv4) with the custom parameter values: min intron length = 5, max intron length = 60000, segment length = 25, segment mismatch = 1 and max multihits = 1. Remaining parameters were kept to default. 

\subsection{Feature extraction}

For extraction of features from the RNA-Seq derived genomic sequences, the transcript sequences were first extracted from the Binary Alignment Map (BAM) file produced by the Tophat2 mapper \cite{Kim2013}. Based on reference alignment of sample reads, a consensus FASTA sequence for each transcript coordinate was constructed by a two-step process: (1) SNP and INDEL calling of BAM file using SAMtools mpileup \cite{Li09} that generated a Variant Call Format (VCF) file, and (2) sequence extraction from the genome and consensus sequence generation using variants from VCF by the SAMtools \textit{faidx} tool \cite{Li09}. The mpileup function collects the information from the BAM file and computes the likelihood. This is stored in a Binary VCF (BCF) format. The Bcftools consensus function creates a consensus FASTA transcript sequence based on reference genome by applying the VCF variants. The sequence obtained can be used for extraction of features for lncRNA classification.

To construct a random forest model, 73 ORF-based and codon-bias features were extracted for each sequence in a given dataset. The features were selected based on the published results on sequence measures and codon bias measures \cite{Fickett1992,Roth2012}. Features extracted from FASTA sequences can be categorized into either ORF-based features or codon bias features. These features constitute a feature set \(F={f_1, f_2, f_3, \ldots, f_{73}}\), where \(f_i\) denotes the \(i\)\textsuperscript{th} feature. The features are discussed below.

\subsubsection{ORF and Sequence based features}

We extracted three ORF-based features: maximum ORF length (\(f_1\)), ORF coverage (\(f_2\)) and mean ORF coverage (\(f_3\)) and four sequence-based features: transcript length (\(f_4\)), GC content (\(f_5\)), Fickett score (\(f_6\)) and Hexamer score (\(f_7\)).

\begin{enumerate}
\item ORF length (\(f_1\)): \(f_1\) is one of the most fundamental features used to distinguish lncRNA from mRNA as majority of protein-coding genes have ORFs greater than 100 amino acids \cite{Frith06}.
\item ORF coverage (\(f_2\)):  \(f_2\) is the ORF coverage defined as length of the longest ORF divided by transcript length. This feature has also been shown to produce good classification performance when compared to ORF length \cite{Wang13,JZhao16}.
\item Overall ORF coverage (\(f_3\)):  \(f_3\) is the overall ORF coverage defined as the average of total ORF lengths divided by transcript length for the sequence.
\item Transcript length (\(f_4\)):  \(f_4\) is the total length of each transcript sequence.
\item GC content (\(f_5\)):  \(f_5\) is the GC content, which is also a common measure to differentiate lncRNA from protein-coding transcripts, as coding sequences have been reported to have higher GC content in exons over introns \cite{Amit12}. The GC content was calculated by counting the frequency of GC motifs for each sequence.
\item Fickett score (\(f_6\)): \(f_6\) is the Fickett score \cite{Fickett82} which was obtained by calculating four base pair position values in the transcript sequence. \(f_6\) is calculated as follows. Let

\(A_{1}\) = Number of A's in positions 1, 4, 7, 10, \ldots,

\(A_{2}\) = Number of A's in positions 2, 5, 8, 11, \ldots, and

\(A_{3}\) = Number of A's in positions 3, 6, 9, 12, \ldots.

Then \(A_{position}\) is defined as,

\begin{equation}
A_{position} = \frac{MAX(A_1,A_2,A_3)}{MIN(A_1,A_2,A_3)+1}
\end{equation}

and \(T_{position}\), \(G_{position}\) and \(C_{position}\) are calculated similarly. In a similar manner, \(A_{content}\), \(T_{content}\), \(G_{content}\) and \(C_{content}\) of the sequence are determined by calculating percentage composition of each base in the sequence. These eight values are then converted to a probability value (\(p\)) using a lookup table \cite{Fickett82} and multiplied by a weight (\(w\)) for each base. The Fickett score \(f_6\) is then determined as:

\begin{equation}
f_6 = \sum_{i=1}^{8}p_{i}w_{i}.
\end{equation}
\item Hexamer score (\(f_7\)): \(f_7\) is the Hexamer score which is computed by making a hexamer table of 4096 ($64\times64$) \textit{k}-mers using a reference set of coding and non-coding sequences. The Hexamer score is calculated by first measuring frequencies of hexamers in the test set sequences. The logarithmic ratio of coding and non-coding sequences is then computed for each hexamer having non-zero frequency in the test set. Positive \(f_7\) indicates higher probability of protein-coding sequence whereas a negative score indicates a higher probability of non-coding RNA sequence. The in-frame hexamer frequency of protein-coding sequences is given by F(\(h_i\)) where i = 1, 2, \ldots, 4096 and in-frame hexamer frequency of lncRNA sequences is given by F'(\(h_i\)) where i = 1, 2, \ldots, 4096. Therefore, for each hexamer sequence, \(H_1\), \(H_1\), \(H_1\), \ldots, \(H_m\), where \(m\) is observed in the test sequence. \(f_7\) is given by:

\begin{equation}
f_7 = \frac{1}{m}\sum_{i=1}^{m}\log\left(\frac{F(h_i)}{F'(h_i).}\right)
\end{equation}
\end{enumerate}

\subsubsection{Codon Bias features}

In protein-coding genes, the translational mapping process of codons (or nucleotide triplets) to amino acids involve usage of synonymous codons which code the same amino acids that are non-distinguishable at protein level. However, it has been reported that there exists a non-uniform codon usage in most genes i.e. codon bias \cite{Clarke70,Ikemura82}. Many indices have been proposed for measuring codon bias, among which we carefully selected six codon-bias measures which are important in distinguishing lncRNAs from mRNAs.

\begin{enumerate}
\item Frequency of the optimal codons (\(f_8\)): This feature is calculated as ratio of the total number of optimal codons to the total number of synonymous codons. The Frequency of the optimal codons (Fop) was also one of the measures proposed by Ikemura \cite{Ikemura82}. The number of codons of optimal codons is calculated as:
\(O_{opt} = \sum_{c\in{\textit{C}_{opt}}} O_{c}\), where \(C_{opt}\) is defined as subset of optimal codons from all codons \(C\) and \(O_{tot}\) is the total number of codons in the sequence. Therefore, \(f_8\) is calculated as: \(f_8 = \frac{O_{opt}}{O_{tot}}\).
\item Codon Usage Bias (\(f_9\)): The Codon Usage Bias (CUB) which assesses codon bias in test set sequence relative to reference set of sequences based on weighted sum of distances of relative codon usage frequencies between the reference set and test set sequences \cite{Karlin96}. The reference set is used as standard to which other sequences can be compared. \(f_9\) is defined as: \(f_9= \sum_{a\in{A}}F_{a}d(f_{a},f_{a}^{ref})\), where \(F_{a}\) is frequency of amino acid \(a\) in the test set sequence; \(f_{a}\) and \(f_{a}^{ref}\) are codon frequencies for amino acid a in test and reference sets, respectively; and d is the L1 norm or manhattan distance for the codon frequency vectors \(f_{a}\) and \(f_{a}^{ref}\). These are calculated as: \(d(f_{a},f_{a}^{ref})=\sum_{c\in{C_{a}}}\mid{f_{ac},f_{a}^{ref}}\mid{}\), where \(f_{ac}\) is the frequency of codon \(c\) encoding amino acid \(a\) in test set sequences and \(f_{a}^{ref}\) is the frequency of amino acid \(a\) in reference set sequences.
\item Relative Codon Bias (\(f_{10}\)): The Relative Codon Bias (RCB) \cite{Roymondal09} is a measure that defines the contribution of codons as: \(w_{c}^{RCB}=\frac{(O_{c}-E[O_{c}])}{E[O_{c}]}\), 
where \(E[O_{c}]\) is the expected number of codon occurrences in three codon positions. Once \(w_{c}^{RCB}\) is determined, \(f_{10}\) is calculated by the following formula for each sequence: 
\(f_{10} =exp\left(\frac{1}{O_{tot}}\sum_{c\in{C}} \log{w_{c}^{RCB}}\right)-1.\) 
\item Weighted sum of relative entropy (\(f_{11}\)): This measures the degree of deviation from equal codon usage \cite{Suzuki04}. Therefore, \(f_{11}\) is defined as sum of relative entropy of each amino acid weighted by its relative frequency in the test sequence which is given by:
\(f_{11} = \sum_{a\in{A}}F_{a}E_{a}\). Here \(F_{a}\) is the relative frequency of amino acid \(a\) in the test sequence and \(E_{a}\) is computed as: 
\(E_{a} = \left(\frac{H_{a}}{\log_2{k_{a}}}\right)\), where \(k_{a}\) number of synonymous codons observed in the test sequence and \(H_{a}\) is the entropy which measures uncertainty of codon usage in the test sequence for amino acid \(a\) computed as:
\(H_{a} =\sum_{c\in{C_{a}}}f_{ac}\log_2{f_{ac}}.\) 
\item Synonymous Codon Usage Order (\(f_{12}\)): This is also an entropy-based codon bias measure and is similar to \(f_{11}\) which differs only by the way entropy is calculated for each amino acid \cite{Wan04}. Instead of calculating the relative entropy, the normalized difference between maximum and observed entropy is computed as
\(E_{a} = \frac{\log_2{k_{a}}-H_{a}}{\log_2{k_{a}}},\) and then \(f_{12}\) is computed as 
\(f_{12} =\sum_{a\in{A}}F_{a}E_{a}.\)
\item Relative Synonymous Codon Usage (RSCU): This measure defines the relationship between observed codon frequencies and the number of times codon is observed when synonymous codon usage is random with no codon bias \cite{Sharp86}. This is calculated as:
\(RSCU_{ac} = \frac{O_{ac}}{\frac{1}{k_a}\sum_{c\in{C_{a}}}O_{ac}}\) where \(O_{ac}\) is the frequency of codon \(c\) for amino acid \(a\). \(RSCU_{ac}\) is the RSCU score (RSCU) for each codon \(c\) encoding amino acid \(a\) and is computed for 61 codons individually by the above equation. Methionine (M), Tryptophan (W) and stop codons were excluded from the analysis as M and W do not have any synonymous codons and stop codons do not contribute any information. Therefore, in total RSCU provided 61 features for the classification: \(f_{13}\), \ldots, \(f_{73}\)  
\end{enumerate}

The random forest model for unified 6-plant species was constructed by building a training set of 22471 balanced coding (positive) and non-coding (negative) sequences, whereas the test set sequences consisted of 7529 sequences. The training and test set sequences in plant RNA-Seq datasets were divided into a proportion containing 75\% and 25\% of the total coding and non-coding sequences. Training and test datasets were constructed as randomized set of 8962 (Training) and 994 (Test) sequences for \textit{A. thaliana}, 5050 (Training) and 562 (Test) sequences for \textit{G. max}, 4968 (Training) and 550 (Test) sequences for \textit{O. sativa}, 7756 (Training) and 860 (Test) sequences for \textit{S. lycopersicum}, 4422 (Training) and 490 (Test) sequences for \textit{S. bicolor}, 7236 (Training) and 802 (Test) sequences for \textit{V. vinifera}, 16902 (Training) and 1878 (Test) sequences for \textit{Z. mays}. For obtaining an optimal set of features, a unified set of 6 plant species, namely, \textit{A. thaliana}, \textit{Z. mays}, \textit{O. sativa}, \textit{B. napus}, \textit{B. rapa} and \textit{B. oleracea}, has been constructed. The dataset has been divided into training and validation sets which were used by LiRFFS algorithm for obtaining an optimal feature set.

\subsection{Feature selection}

The selection of optimal features is an important optimization for classification. Wrapper-based Feature Selection (FS) methods such as Sequential Forward Selection (SFS) \cite{Pudil94} or SVM-recursive feature elimination (SVM-RFE) \cite{Huang14} are computationally inefficient and can fail to identify optimal feature subsets. Whereas filter-based FS methods such as mRMR \cite{Peng05}, Chi-square \cite{Chen11} and Information Gain \cite{Lee06}, assign relevance score or rank to each feature by considering each feature separately and ignoring dependencies between features which leads to worse classification performance.  Regression based approaches, such as least-squares estimate methods, often produce larger variance during model fitting which leads to over-fitting and poor generalization. Least Absolute Shrinkage and Selection Operator (LASSO) is a feature selection method which combines least-square loss with L1 norm constraint and produces sparse features by shrinking coefficients to zero. Other approaches such as ridge regression \cite{Tibshirani96,Marquardt70} use L2 norm due to which it produces non-zero coefficients and therefore becomes inefficient for feature selection. Usage of Lq norm (with q $<$ 1 or q $>$ 1) approaches for optimization are generally non-convex and make the minimization computationally challenging.

\begin{figure}[h]
\centerline{\includegraphics[width=\linewidth]{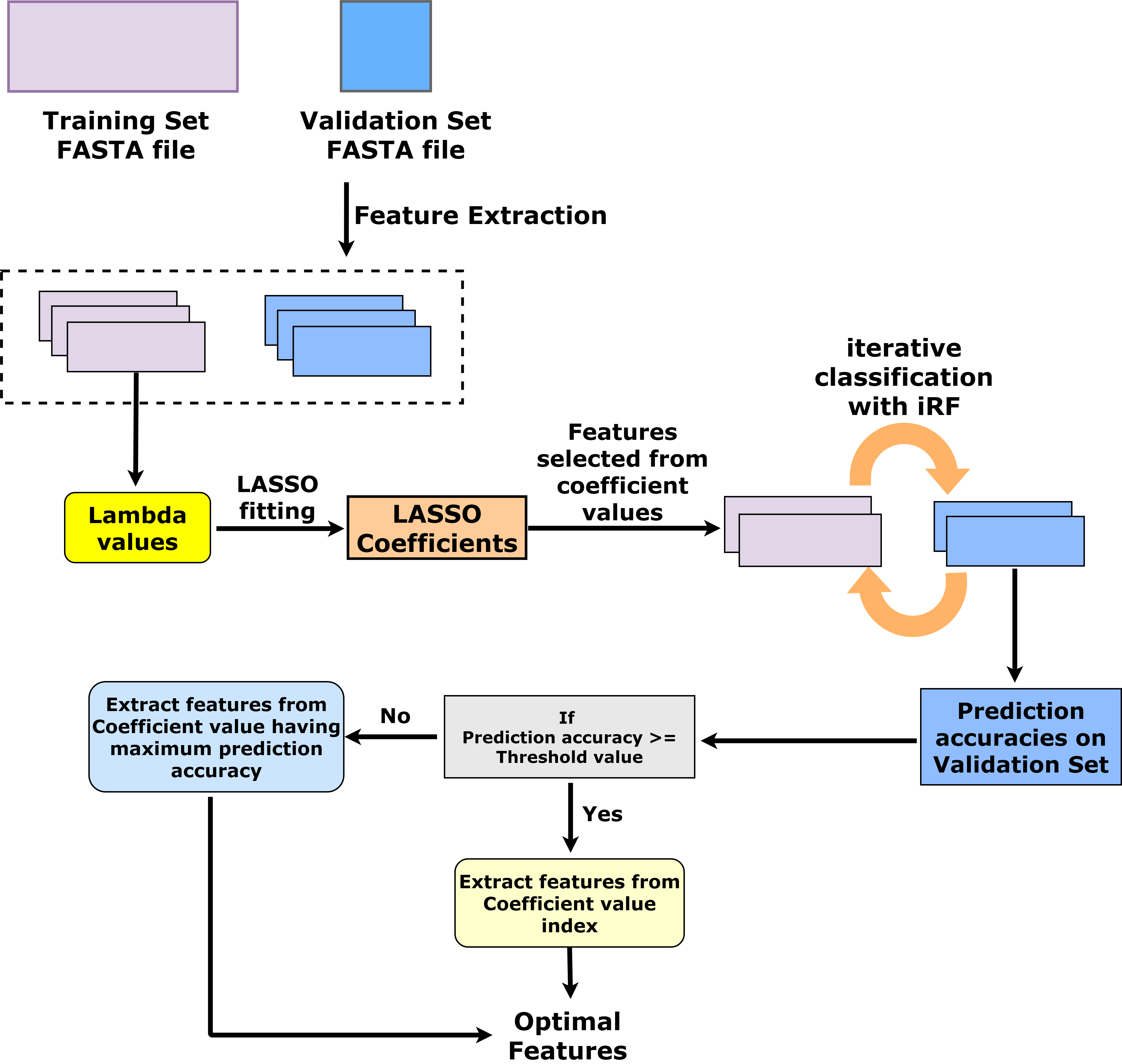}}
\caption{LiRFFS workflow. Sequence and codon-bias features from the training and validation lncRNA and protein-coding sequences are extracted. LASSO coefficients are generated from the training set and iteratively applied on the validation set using an iRF classifier to generate the prediction accuracy at each $\lambda$ value. Prediction accuracies are compared against the minimum threshold tolerance value. If the prediction accuracy produced by a particular $\lambda$ value is $\geq$ minimum threshold tolerance value, the optimal features are selected from the filtered coefficient set. If accuracy is lower then the $\lambda$ value, the optimal features are selected from the $\lambda$ value producing the maximum prediction accuracy.}\label{fig:01}
\end{figure}

The PLIT framework implements LASSO and an iterative Random Forest Feature Selection (LiRFFS) algorithm (Algorithm 1) for identifying the principal set of collective features yielding the highest accuracy. It works by iterative selection of features based on varying the value of LASSO parameter lambda ($\lambda$) (Figure~1\vphantom{\ref{fig:01}}).

As $\lambda$ changes, non-zero beta coefficients are generated which corresponds to the selection of features using L1-regularized optimization of LASSO \cite{Tibshirani96}. The $\beta$ coefficients are calculated on training set features for each $\lambda$ using the following equation:

\begin{equation}
\beta^{LASSO} = \argmin_r \frac{1}{2n}{\norm{X\beta{} - y}_{2}^{2}} + \lambda{\norm{\beta{}}_{1}} 
\end{equation}
where, $\lambda$ $\geq$ 0, \({\norm{X\beta{} - y}_{2}^{2}}\) is the loss function (i.e. sum of squares), \(\norm{\beta{}}_{1}\) is the penalty term and $\lambda$ is the tuning parameter which controls the strength of the penalty. Features extracted from the coding and noncoding sequences are divided into training and validation sets. $\beta$ coefficients are calculated on each $\lambda$ value. The selected features for each $\lambda$ are iteratively applied on the validation set to obtain the accuracy vector. The optimal feature set is obtained by selecting the feature set that produces the prediction accuracy between the tolerance accuracy value and the maximum prediction accuracy value. The optimal feature set can be used for building the model for classification of test set transcript sequences. 

The algorithm (Algorithm 1) takes \(\lambda_{lower}\), \(\lambda_{upper}\) and \(\lambda_{step-size}\) as lower limit, upper limit and step-size input values for creation of \(\lambda\) values. The default values of \(\lambda_{lower}\), \(\lambda_{upper}\) and \(\lambda_{step-size}\) are chosen as 0.00001, 0.1 and 0.00001, respectively. ntrees is defined as number of trees required for generating random forests (ntrees=400 (default)). Tolerance is the maximum threshold value allowed (tolerance=0.5). \(X_{train}\), \(X_{val}\), \(Y_{train}\) and \(Y_{val}\) are the features and class values of training and validation sets, respectively. The function \(estimateBetaLASSO\) is used for calculation of \(\beta^{LASSO}\) values which is computed on the value of \(\lambda\). The \(\beta^{LASSO}\) values are calculated by: \(\frac{1}{2n}{\norm{X\beta{} - y}_{2}^{2}} + \lambda{\norm{\beta{}}_{1}}\). Coordinate Descent (CD) minimization algorithm is implemented for minimizing the objective function with respect to each of its coordinate directions. Results of the \(\beta\) coefficient values for each \(\lambda\) value is stored in the betaE list. \(p\) contains the number of features from the training set. selProb contains the selection probabilities of the features for the classification using iterative Random Forests. maxAccIndex contains the maximum prediction accuracy among n iterations of validation set sequences for particular value of \(\lambda\), and accRfPred contains the prediction accuracies of all \(\lambda\) values. The algorithm returns a feature list of minimal and maximal optimal features.  

 \begin{spacing}{0.8}
 \begin{algorithm}
 \caption{LiRFFS algorithm}
 \begin{algorithmic}[1]
 \renewcommand{\algorithmicrequire}{\textbf{Input:}}
 \renewcommand{\algorithmicensure}{\textbf{Output:}}
 \REQUIRE \text{$\lambda_{lower}$, $\lambda_{upper}$, $\lambda_{stepsize}$, $\beta$, n, tolerance, ntrees, \(X_{train}, X_{val}, Y_{train}, Y_{val}\)}
 \ENSURE  \text{Minimal and maximal optimal feature lists}
  \STATE{\(\lambda\) $=$  Create list based on \(\lambda_{lower}\),\(\lambda_{upper}\) and \(\lambda_{step-size}\) values}
  \STATE \text{betaLASSO $=$ function \textit{estimateBetaLASSO}($X_{train}$,$Y_{train}$)}
 \\ 
  \FOR {$i = 0$ to $length(\lambda)$}
  \STATE \text{betaE $= minimize$ the betaLASSO using CD minimization}
  \IF {($values$ in $betaE <$ tolerance)}
  \STATE Set $values$ in betaE $= 0$
  \STATE betaENonZero $= length(values$ in betaE $= 0)$
  \ENDIF
  \STATE \text{betaEArray $=$ betaE}
  \IF {(betaENonZero $< length($betaEArray$-1)$)}
  \FOR{$j = 1$ to betaENonZero}
  \STATE \text{[\(X_{trainF},X_{valF},Y_{trainF},Y_{valF}] = [X_{train}[j],X_{val}[j],Y_{train}[j],X_{val}[j]]\)}
  \ENDFOR
  \STATE \text{selProb $=$ replicate the values of \(\frac{1}{p}\)}
  \STATE \text{Initialise \textit{rf} as \textit{list}}
  \FOR{$iter = 1$ to n}
  \STATE \text{rf[iter] $=$ RandomForest(\(X_{trainF},Y_{trainF},X_{valF},Y_{valF}\),selProb,ntrees)}
  \STATE \text{selProb $=$ \textit{giniImportance}(rf[iter])}
  \ENDFOR
  \STATE \text{maxAccIndex $=$ extract index values of the \(\lambda\) value having}
 \text{the maximum prediction accuracy stored in \(rf[iter]\)}
  \STATE \text{Store the maxAccIndex value in accRfPred list}
  \ENDIF
  \ENDFOR
  \FOR{$i = $maxAccIndex to 0}
  \STATE \text{diffArrNeg $=$ accRfPred[$i$] - accRfPred[$i-1$]}
  \ENDFOR
  \FOR{$i = $maxAccIndex to length(accRfPred)}
  \STATE \text{diffArrPos $=$ accRfPred[$i$] - accRfPred[$i+1$]}
  \ENDFOR
  \STATE \text{Extract index values of diffArrNeg and diffArrPos values} 
  \text{and store in thresArrNeg and thresArrPos lists}
  \STATE \text{Extract Last elements from thresArrNeg and thresArrPos lists}
 \RETURN Optimal feature list based on thresArrPos and thresArrNeg lists 
 \end{algorithmic} 
 \end{algorithm}
 \end{spacing}

\subsection{10-fold cross validation and repeated 10-fold cross validation with data shuffling}

For evaluating the prediction accuracy of PLIT against CPC2, CPAT, lncScore and PLEK tools, a 10-fold Cross Validation (CV) and repeated 10-fold CV with data shuffling benchmarking was performed on the coding and non-coding sequences extracted from the RNA-seq datasets. From the complete sequence set, 10\% were selected as test set and 90\% as training set in each fold consisting of balanced lncRNA and protein-coding sequences. For repeated 10-fold CV, five repetitions were performed with shuffling of sequences in each iteration followed by 10-Fold CV in each repetition. 

\subsection{Performance evaluation criteria}

To assess classification performance of lncRNAs and mRNA transcripts, Accuracy (ACC), Sensitivity (SENS), Specificity (SPEC), F1-Score (F1), Negative Predictive Value (NPV) and Matthews Correlation Coefficient (MCC) metrics were used which are defined as follows.

\begin{itemize}
\renewcommand\labelitemi{--}
\item Accuracy \(= \frac{TP + TN}{TP + FP + FN + TN}\) 
\item Sensitivity or Recall \(= \frac{TP}{TP + FN}\)
\item Specificity \(= \frac{TN}{FP + TN}\)
\item Positive Predictive Value or Precision \(= \frac{TP}{TP + FP}\)
\item F1-Score \(= \frac{2 * (Precision * Recall)}{Precision + Recall}\)
\item Negative Predictive Value or NPV \(= \frac{TN}{TN + FN}\)
\item Matthews Correlation Coefficient or MCC \(= \frac{(TP * TN) - (FP * FN)}{\sqrt{(TP + FP)(FN + TN)(FP + TN)(TP+FN)}}\)
\end{itemize}
In all the above, TP = True Positive, TN = True Negative, FP = False Positive and FN = False Negative.

\section{Results}

\subsection{Results of optimal feature selection}

\begin{figure}[h]
\centerline{\includegraphics[width=\linewidth]{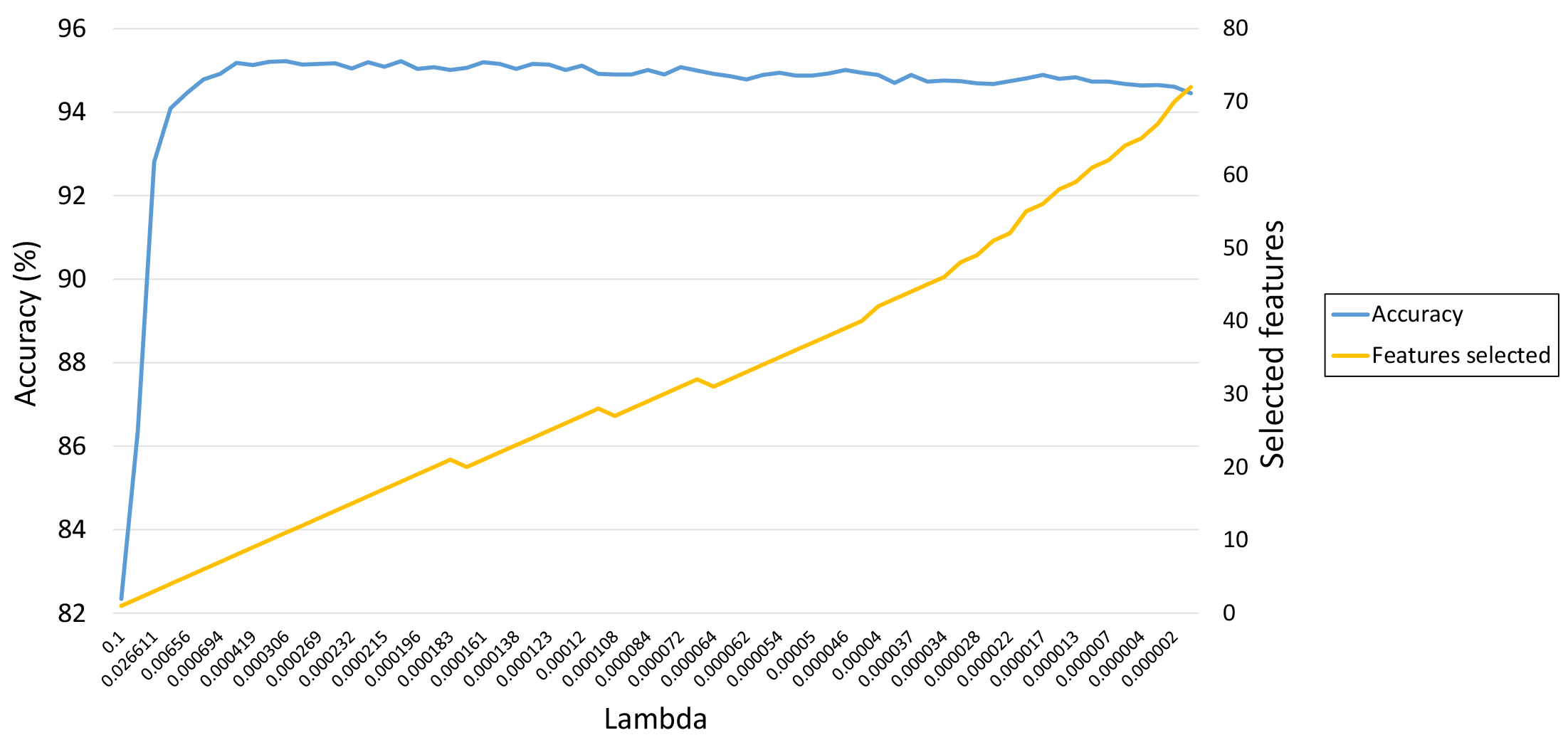}}
\caption{Feature selection bar plot. The bar plot shows prediction accuracy of the features selected at different $\lambda$ values performed on validation set lncRNA and protein-coding transcripts on six plant species. X-axis shows range of $\lambda$ values ranging from 0.1 to \(1 \times 10^{-6}\). Primary y-axis shows the prediction accuracy and secondary y-axis shows the number of features selected at each $\lambda$ value.}\label{fig:02}
\end{figure}

The selection of optimal features was performed on a unified dataset of 6 plant species (\textit{A. thaliana}, \textit{Z. mays}, \textit{O. sativa}, \textit{B. napus}, \textit{B. rapa} and \textit{B. oleracea}). The dataset consisted of 22,468 (lncRNA and mRNA) transcript sequences selected as training set and 7,532 sequences selected as validation set. An optimal feature set was selected based on \(\lambda\) values ranging from 0.1 to $1.0\times 10^{-5}$. Figure~2\vphantom{\ref{fig:02}} shows the prediction accuracies of lncRNA and mRNA sequences in the validation set over the range of \(\lambda\) values. Based on \textit{tolerance} cutoff value of 0.5, two feature sets, namely, the 7 feature set (7F) and the 31 feature set (31F) were selected having minimal and maximal optimal features. 7F is selected based on the least number of features producing higher prediction accuracy having accuracy within the \textit{tolerance} threshold value from the maximum prediction accuracy \(\lambda\) value. Whereas 31F is selected based on the maximum number of features having prediction accuracy within the \textit{tolerance} threshold value from the maximum prediction accuracy \(\lambda\) value. Prediction of test set sequences was performed based on the optimal feature sets obtained from LASSO-iRF computation. 

The performance of the 7F, 31F and 73F feature sets were compared on \textit{Arabidopsis thaliana}, \textit{Zea mays}, \textit{Sorghum bicolor} and \textit{Vitis vinifera} RNA-seq test sets containing lncRNA and mRNA transcripts. The AUC plots (Figure~3\vphantom{\ref{fig:03}}) shows that performance of 31F was similar to 73F, whereas 7F showed slight decrease in the AUC value as compared to 31F and 73F. A difference of 0.02, 0.0085 and 0.02 in the AUC values for 7F can be observed in \textit{Arabidopsis thaliana}, \textit{Sorghum bicolor} and \textit{Vitis vinifera} test set sequences (Figure~3\vphantom{\ref{fig:03}}(a), (c) and (d)). Although the difference is minor, prediction of lncRNAs by the 7F feature set increases the False Positive Rate as compared to the 31F feature set.           

\begin{figure}
\begin{subfigure}{.5\textwidth}
  \centering
  \includegraphics[width=1\linewidth]{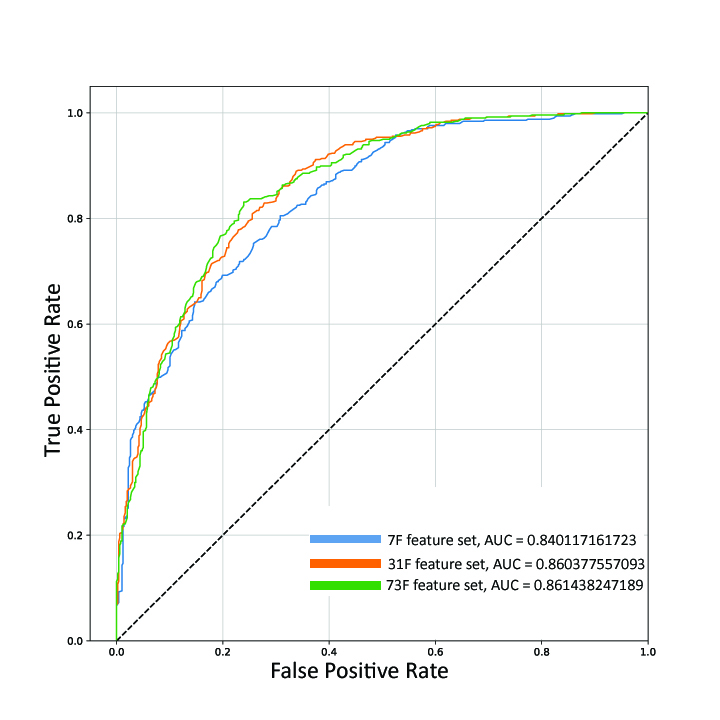}
  \centerline{(a)}
  \label{fig:sfig1}
\end{subfigure}%
\begin{subfigure}{.5\textwidth}
  \centering
  \includegraphics[width=1\linewidth]{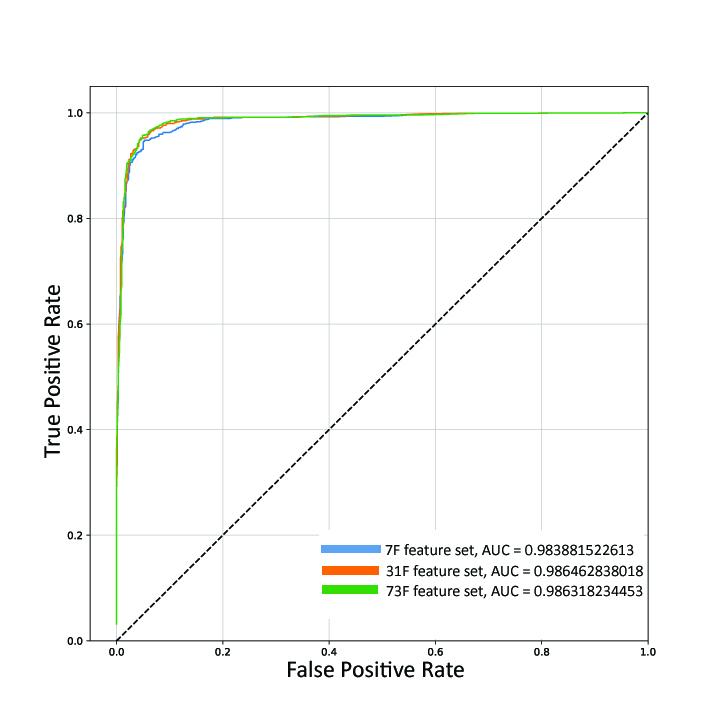}
  \centerline{(b)}
  \label{fig:sfig1}
\end{subfigure}
\begin{subfigure}{.5\textwidth}
  \centering
  \includegraphics[width=1\linewidth]{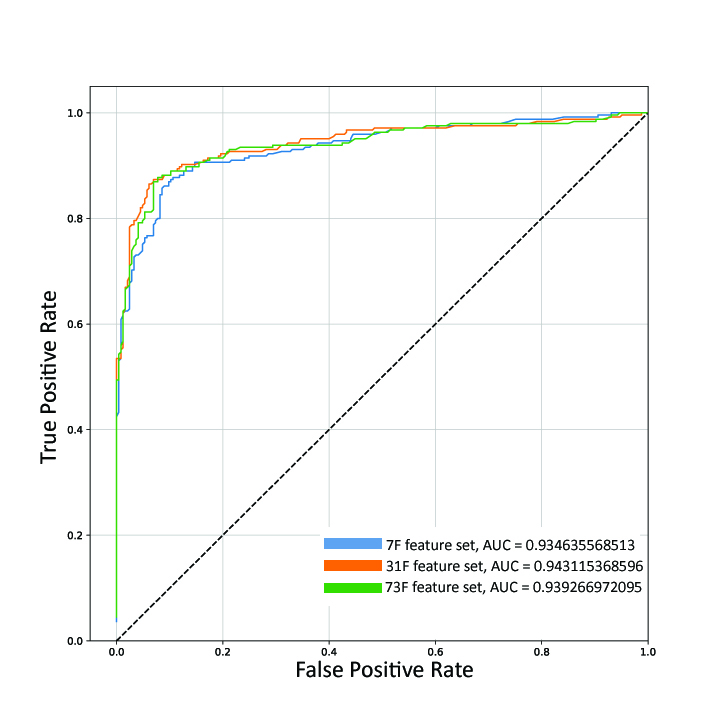}
  \centerline{(c)}
  \label{fig:sfig1}
\end{subfigure}
\begin{subfigure}{.5\textwidth}
  \centering
  \includegraphics[width=1\linewidth]{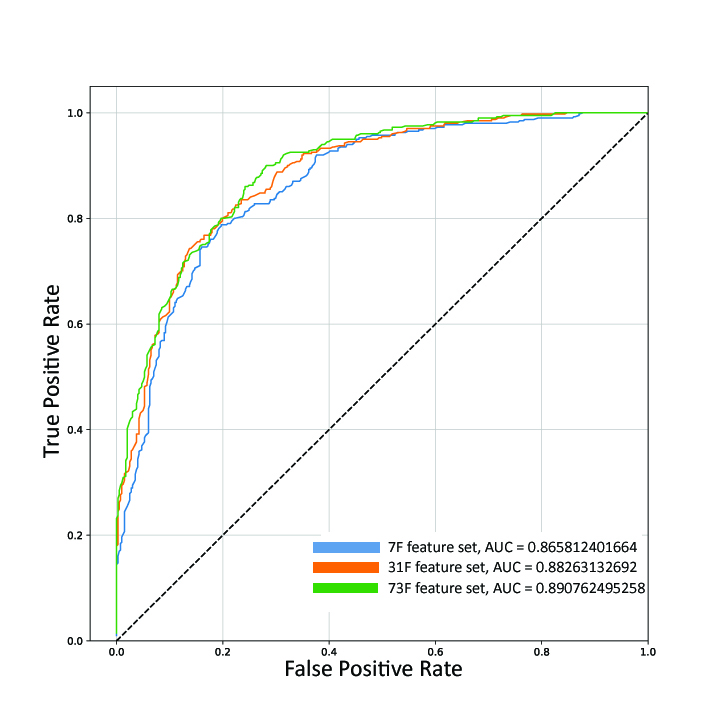}
  \centerline{(d)}
  \label{fig:sfig1}
\end{subfigure}
\caption{ROC plots of (a) \textit{Arabidopsis thaliana}, (b) \textit{Zea mays}, (c) \textit{Sorghum bicolor}, and (d) \textit{Vitis vinifera} RNA-seq test datasets, showing comparison of AUC values for 7F, 31F and 73F feature sets.}
\label{fig:fig}
\end{figure}

\subsection{Performance of PLIT feature groups on plant RNA-seq datasets}

To evaluate performance of PLIT, the tool was evaluated on coding and non-coding sequences obtained from seven plant RNA-seq species. Individual performance of PLIT in different species (Table~2\vphantom{\ref{Tab:02}}) shows that on an average, a higher accuracy and specificity values were obtained for \textit{A. thaliana}, \textit{G. max }, \textit{O. sativa}, \textit{S. lycopersicum}, \textit{V. vinifera} and \textit{Z. mays} datasets. High specificity value indicate accuracy in identification of lncRNA sequences in the RNA-seq dataset. An evaluation of F1 and accuracy values displayed precision in detecting lncRNA sequences. A measure of NPV also demonstrates accuracy in determining the true positives and true negatives from the test set.     

\begin{table}[h!]
\vspace{-0.2cm}
\centering
\caption{Performance (percentage accuracy) of PLIT on different plant RNA-seq datasets.}
\begin{tabular}{l@{\hspace{0.3cm}}l@{\hspace{0.6cm}}l@{\hspace{0.4cm}}l@{\hspace{0.4cm}}l@{\hspace{0.4cm}}l@{\hspace{0.4cm}}l}
  \hline
Data set & ACC & SENS & SPEC & F1 & NPV & MCC\\
  \hline
\textit{Arabidopsis thaliana} & 77.16 & 76.05 & 78.27 & 77.16 & 76.57 & 0.54\\
\textit{Glycine max} & 84.16 & 82.92 & 85.41 & 84.16 & 83.33 & 0.68\\
\textit{Oryza sativa} & 87.09 & 86.18 & 88.00 & 87.09 & 86.42 & 0.74\\ 
\textit{Solanum lycopersicum} & 85.35 & 83.95 & 86.74 & 85.34 & 84.39 & 0.70\\
\textit{Sorghum bicolor} & 88.57 & 88.98 & 88.16 & 88.57 & 88.88 & 0.77\\
\textit{Vitis vinifera} & 81.04 & 75.81 & 86.28 & 81.00 & 78.10 & 0.62\\
\textit{Zea mays} & 94.94 & 94.25 & 95.63 & 94.94 & 94.32 & 0.89\\
 \hline
\end{tabular}
\label{Tab:01}
\end{table} 

An area under receiver operating characteristic (AUC) curve gives better insight about the ability of a classifier to separate two classes. From the RNA-seq datasets, a higher AUC was observed for \textit{S. lycopersicum}, \text{G. max}, \textit{O. sativa}, \textit{V. vinifera}, and \textit{Z. mays} with an average AUC of 0.933 (Figure~4\vphantom{\ref{fig:04}}). A slightly lower AUC was observed for \textit{S. bicolor} and \textit{A. thaliana} having AUC of 0.75 and 0.85 respectively. The AUC and the evaluation metrics clearly indicates that PLIT predict the lncRNA sequences in plants without overfitting the training data. 

The performance of PLIT was benchmarked against the four popular coding potential computation tools: CPAT, CPC2, PLEK and lncScore, using the plant RNA-seq test datasets. An initial benchmarking analysis based on the prediction accuracy (Table~3\vphantom{\ref{Tab:03}}) shows that PLIT exhibited much higher prediction accuracies in all the plant RNA-seq test set sequences. The accuracy of PLIT ranged from 76.5\% to 96.7\%, while only lncScore achieved the accuracy >90\% among other tools. The accuracy for lncScore ranged between 62.7\% to 92.6\%, whereas CPC2, PLEK and CPAT demonstrated accuracies between 52.1\% and 89.07\%. Apart from accuracy, it was noticed that PLIT achieved higher sensitivity and specificity in all the test sets which indicates that PLIT is higher quality classifier for plants. CPAT demonstrated comparable sensitivity but demonstrated much lower specificity. On the other hand, CPC2, PLEK and lncScore achieved higher specificity but much lower sensitivity. Lower sensitivity implies producing higher false negatives i.e. classifying coding sequences as long non-coding transcripts, whereas lower specificity implies increase in false positive results i.e. classifying long non-coding as coding transcripts (Table~4\vphantom{\ref{Tab:04}} and Table~5\vphantom{\ref{Tab:05}}). Table 4 and 5 demonstrates that CPC2 and lncScore shows lower Sensitivity and higher Specificity values for all the plant species, whereas CPAT and PLEK produces lower Specificity and higher Sensitivity values. Overall, the results demonstrated PLIT as more accurate and robust tool for differentiating lncRNA and protein-coding transcripts in plants exhibiting consistently greater accuracy, Sensitivity and Specificity metrics in all plant species.      



\begin{figure}[h]
\centerline{\includegraphics[width=\linewidth]{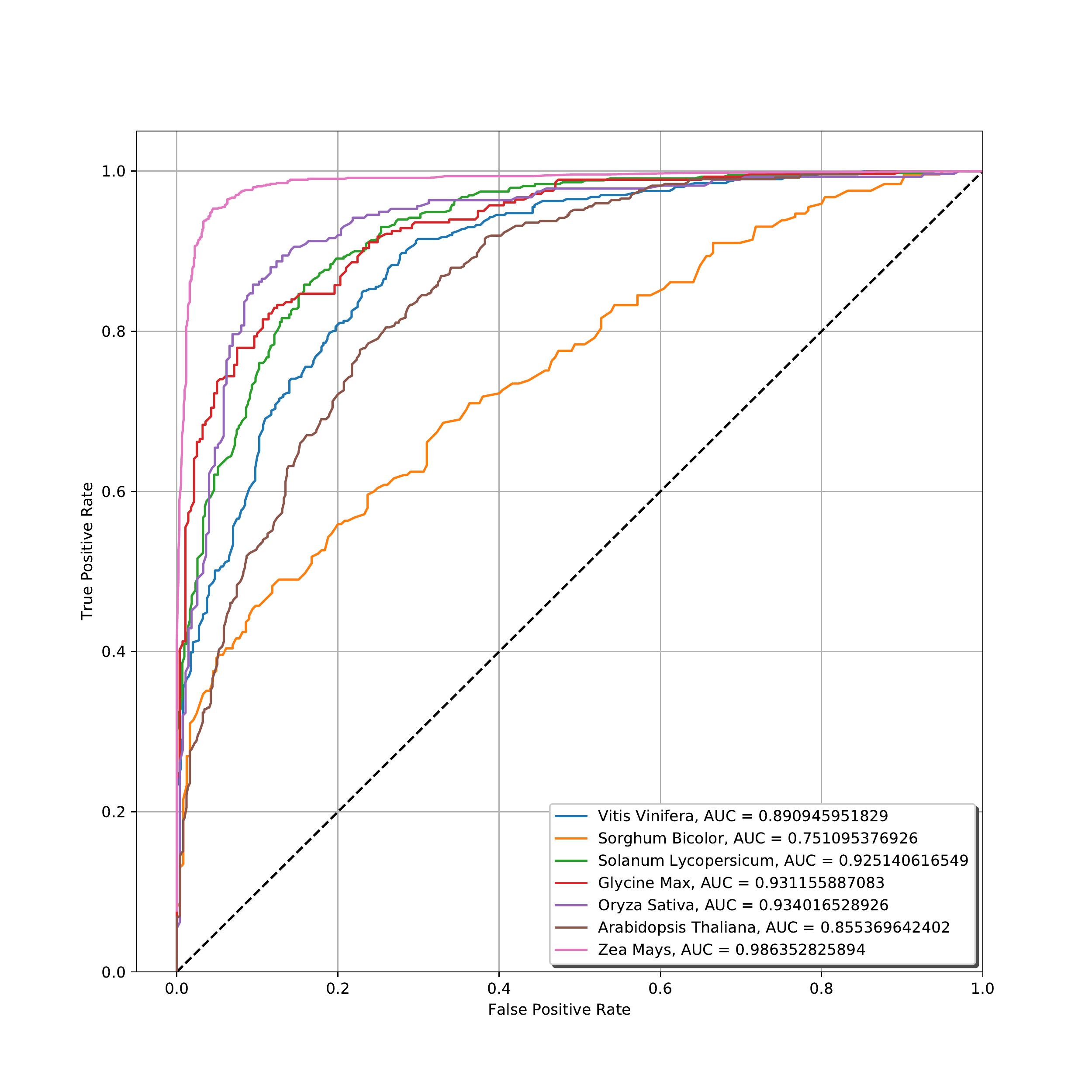}}
\caption{ROC curves and AUC values of PLIT for different RNA-seq plant data sets.}\label{fig:04}
\end{figure}

\begin{table}[h!]
\vspace{-0.2cm}
\centering
\caption{Performance comparison (percentage accuracy) of PLIT with existing tools on different plant RNA-seq datasets.}
\begin{tabular}{l@{\hspace{0.3cm}}l@{\hspace{0.6cm}}l@{\hspace{0.4cm}}l@{\hspace{0.4cm}}l@{\hspace{0.4cm}}l@{\hspace{0.4cm}}l}
  \hline
Data set & CPAT & CPC2 & PLEK & lncScore & PLIT\\
  \hline
\textit{Arabidopsis thaliana} & 53.01 & 50.40 & 64.98 & 67.70 & \textbf{76.05}\\
\textit{Glycine max} & 65.42 & 57.98 & 67.90 & 73.22 & \textbf{86.88}\\
\textit{Oryza sativa} & 64.49 & 66.30 & 69.38 & 78.44 & \textbf{86.59}\\ 
\textit{Solanum lycopersicum} & 55.58 & 58.48 & 63.02 & 66.74 & \textbf{86.04}\\
\textit{Sorghum bicolor} & 66.93 & 60.41 & 64.69 & 71.22 & \textbf{87.96}\\
\textit{Vitis vinifera} & 52.24 & 58.97 & 65.21 & 63.34 & \textbf{78.30}\\
\textit{Zea mays} & 87.91 & 59.53 & 79.39 & 92.33 & \textbf{96.43}\\
 \hline
\end{tabular}
\label{Tab:01}
\end{table} 

\begin{table}[h!]
\vspace{-0.2cm}
\centering
\caption{Sensitivity comparison of PLIT with existing tools on different plant RNA-seq datasets.}
\begin{tabular}{l@{\hspace{0.3cm}}l@{\hspace{0.6cm}}l@{\hspace{0.4cm}}l@{\hspace{0.4cm}}l@{\hspace{0.4cm}}l@{\hspace{0.4cm}}l}
  \hline
Data set & CPAT & CPC2 & PLEK & lncScore & PLIT\\
  \hline
\textit{Arabidopsis thaliana} & 65.30 & 32 & 67.75 & 57.96 & 76.25\\
\textit{Glycine max} & 82.70 & 32.62 & 57.76 & 66.06 & 86.17\\
\textit{Oryza sativa} & 78.17 & 47.10 & 67.27 & 77.81 & 88.40\\ 
\textit{Solanum lycopersicum} & 72.22 & 27.44 & 56.65 & 63.19 & 84.18\\
\textit{Sorghum bicolor} & 81.68 & 48.16 & 73.06 & 68.16 & 88.98\\
\textit{Vitis vinifera} & 80 & 35.91 & 61.30 & 56.53 & 73.81\\
\textit{Zea mays} & 91.39 & 44.51 & 87.64 & 87.22 & 96,16\\
 \hline
\end{tabular}
\label{Tab:01}
\end{table}

\begin{table}[h!]
\vspace{-0.2cm}
\centering
\caption{Specificity comparison of PLIT with existing tools on different plant RNA-seq datasets.}
\begin{tabular}{l@{\hspace{0.3cm}}l@{\hspace{0.6cm}}l@{\hspace{0.4cm}}l@{\hspace{0.4cm}}l@{\hspace{0.4cm}}l@{\hspace{0.4cm}}l}
  \hline
Data set & CPAT & CPC2 & PLEK & lncScore & PLIT\\
  \hline
\textit{Arabidopsis thaliana} & 51.67 & 68.81 & 63.18 & 78.27 & 75.85\\
\textit{Glycine max} & 60.09 & 83.33 & 79.07 & 81.56 & 87.58\\
\textit{Oryza sativa} & 59.75 & 85.50 & 71.74 & 79.34 & 84.78\\ 
\textit{Solanum lycopersicum} & 53.19 & 89.53 & 71.62 & 72.79 & 87.90\\
\textit{Sorghum bicolor} & 61.56 & 72.65 & 52.65 & 77.55 & 86.93\\
\textit{Vitis vinifera} & 51.16 & 82.04 & 69.57 & 70.57 & 82.79\\
\textit{Zea mays} & 84.97 & 74.54 & 71.14 & 97.44 & 96.70\\
 \hline
\end{tabular}
\label{Tab:01}
\end{table}

\subsection{Results of 10-fold Cross Validation (CV) performance benchmarking}

To assess the performance of PLIT in plant RNA-seq datasets, a 10-fold CV performance benchmarking was performed. The prediction accuracy of PLIT against CPAT, CPC2, PLEK and lncScore was tested on each fold. To evaluate the classification performance of PLIT, 31F was used for comparing the prediction accuracies of test sets.

Transcript length distribution of TAIR10-annotated and EST-derived lncRNA transcripts demonstrates the degree of sequence length variation in lncRNA transcripts (Figure~5\vphantom{\ref{fig:05}}). Sequences derived from the TAIR10 annotation data ranges between 200 bp and 8000 bp whereas sequences derived from EST analysis ranges widely between 200 bp and $7.8\times 10^5$ bp. Additionally, ORF count of EST-lncRNA sequences reveals counts greater than 700 ORFs per frame. Such extremely long lncRNA sequences are generally mis-classified as protein-coding transcripts, due to which the overall prediction accuracy decreases.

\begin{figure}[h]
\centerline{\includegraphics[width=\linewidth]{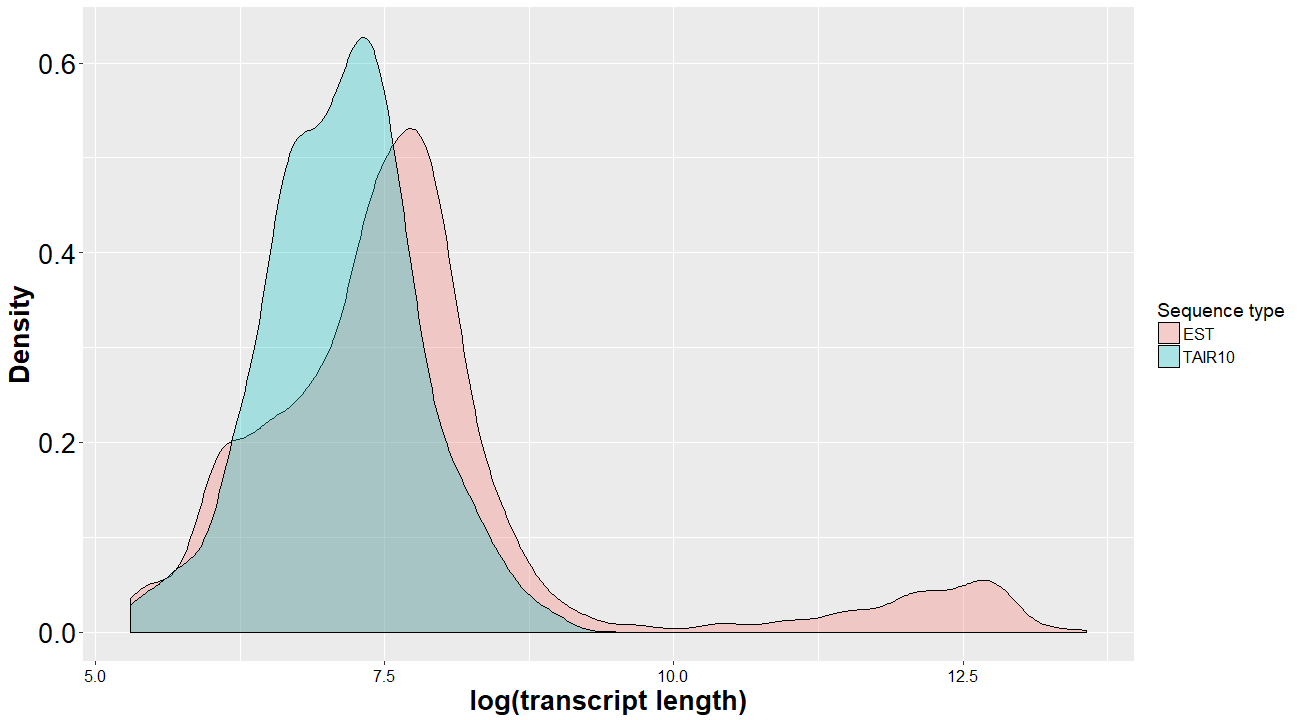}}
\caption{Density distribution of transcript lengths of lncRNA sequences in ATH TAIR10-annotated and EST-predicted results. X-axis is log of transcript lengths and y-axis is density.}\label{fig:04}
\end{figure}

The 10-fold CV benchmarking on seven plant RNA-seq test set sequences (Figure~6\vphantom{\ref{fig:06}}) demonstrates PLIT's superior performance exhibiting significant improvement in the prediction accuracy across all the folds. Among all the tools, CPC2 demonstrated poor accuracy in predicting the lncRNAs in the \textit{A. thaliana}, \textit{G. max}, \textit{O. sativa}, \textit{S. bicolor} and \textit{Z. mays} datasets. The prediction accuracies ranged between 49.7\% and 53\% for \textit{A. thaliana}, 57\% to 63\% for \textit{G. max}, 63\% to 69\% for \textit{O. sativa}, 56\% to 66\% for \textit{S. bicolor} and 50\% to 60\% for \textit{Z. mays}. Similar to CPC2, CPAT also generated lower prediction accuracies with accuracies ranging between 49\% to 54\% for \textit{A. thaliana} and \textit{V. vinifera} species. CPAT generated comparatively higher prediction accuracy against PLEK and CPC2 on \textit{S. bicolor} dataset. PLEK and lncScore exhibited similar performance for \textit{A. thaliana}-EST derived sequences, \textit{G. max} and \textit{V. vinifera} displaying accuracies in the range of 60\% to 76\%. lncScore, on the other hand, demonstrated higher performance among CPAT, PLEK and CPC2 tools with accuracies ranging between 65\% and 70\% for \textit{A. thaliana} TAIR10 annotated lncRNA sequences and \textit{S. lycopersicum}, 74\% to 80\% for \textit{O. sativa}, and 70\% to 78\% for \textit{S. bicolor}. The difference in the accuracies between PLIT and other tools ranges between 9\% to 29\% in \textit{A. thaliana}, 11\% to 27\% for \textit{G. max}, 7\% to 20\% for \textit{O. sativa}, 16\% to 29\% for \textit{S. lycopersicum}, 12\% to 25\% for \textit{S. bicolor}, 13\% to 27\% for \textit{V. vinifera}, and 3\% to 46\% for \textit{Z. mays}. However, for the \textit{Z. mays} test set transcript sequences, the highest difference was produced by CPC2 alone. CPAT and PLEK generated accuracy difference ranging between 3 to 8\%, whereas PLEK exhibited a difference of 16\% against PLIT.  

\begin{figure}
\begin{subfigure}{.5\textwidth}
  \centering
  \includegraphics[width=1.0\linewidth]{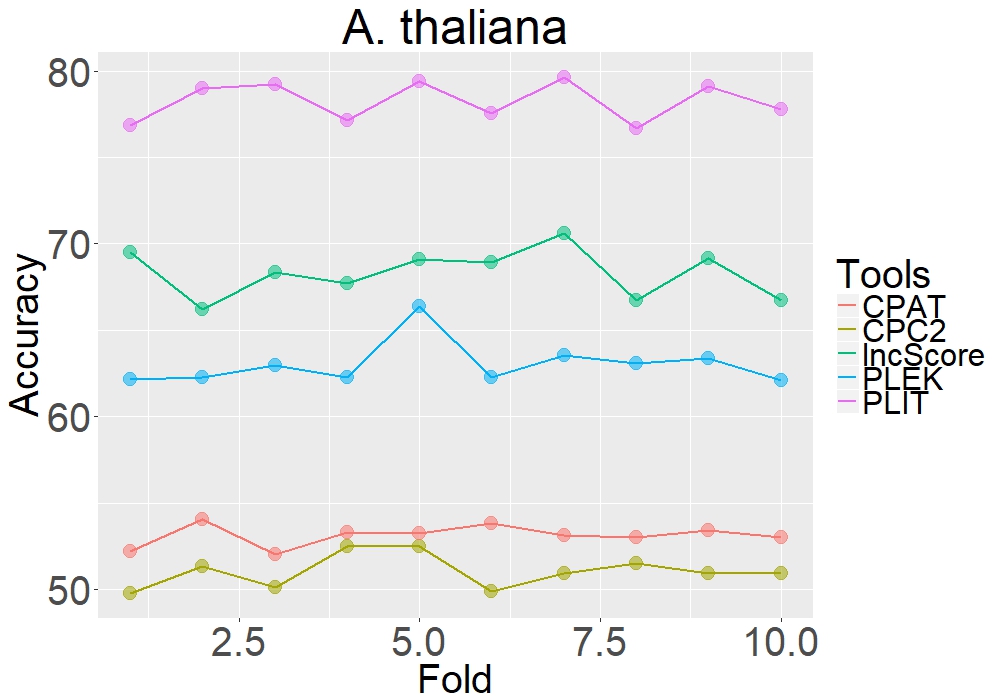}
  \centerline{(a)}
  \label{fig:sfig1}
\end{subfigure}%
\begin{subfigure}{.5\textwidth}
  \centering
  \includegraphics[width=1.0\linewidth]{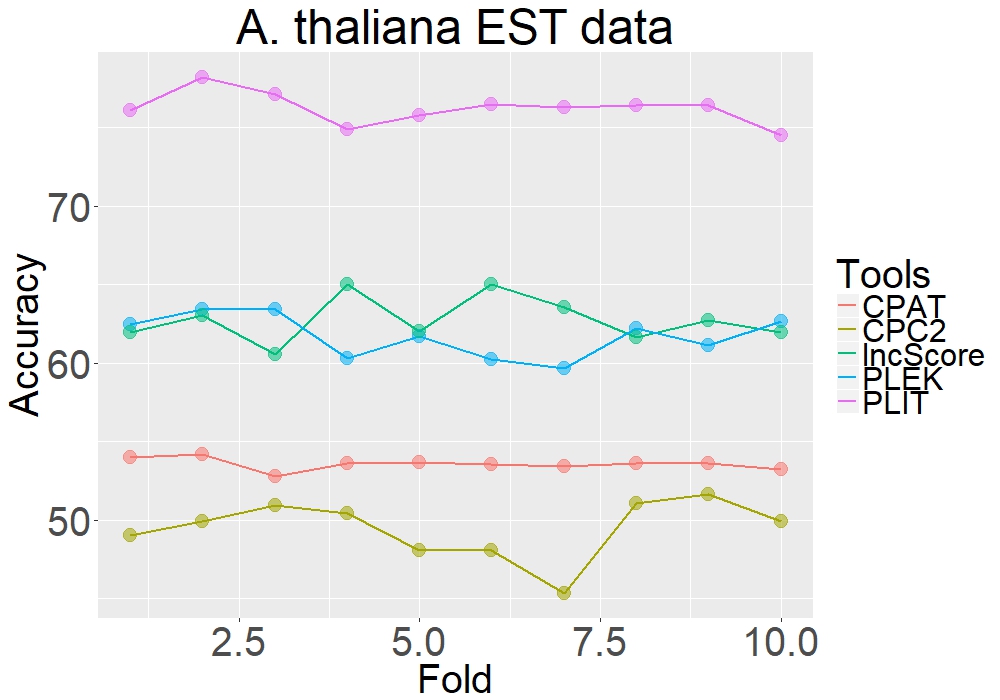}
  \centerline{(b)}
  \label{fig:sfig1}
\end{subfigure}
\begin{subfigure}{.5\textwidth}
  \centering
  \includegraphics[width=1.0\linewidth]{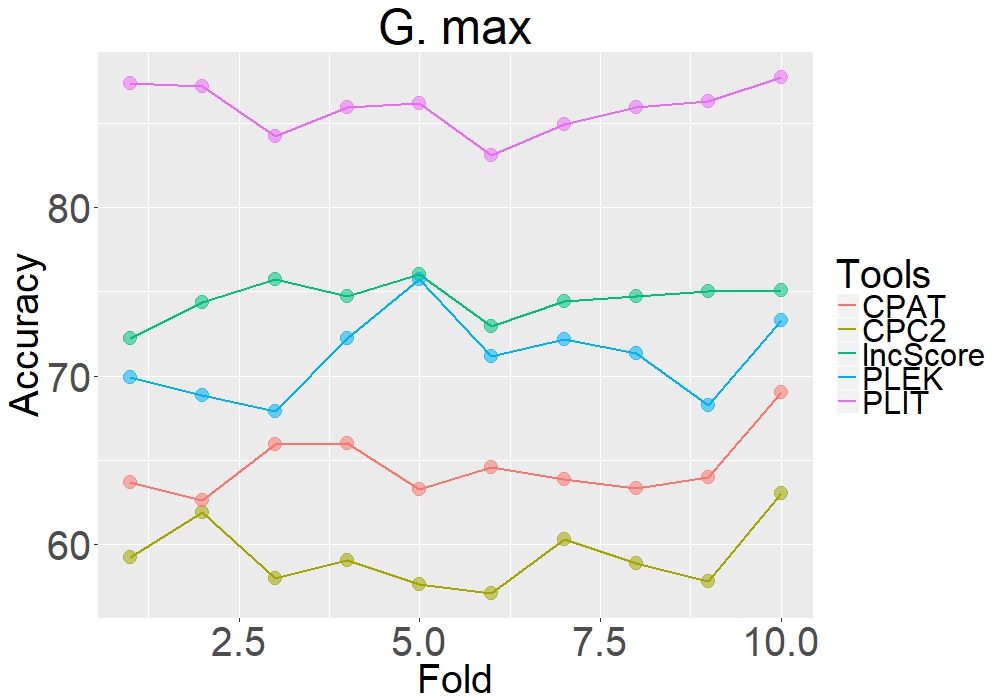}
  \centerline{(c)}
  \label{fig:sfig2}
\end{subfigure}
\begin{subfigure}{.5\textwidth}
  \centering
  \includegraphics[width=1.0\linewidth]{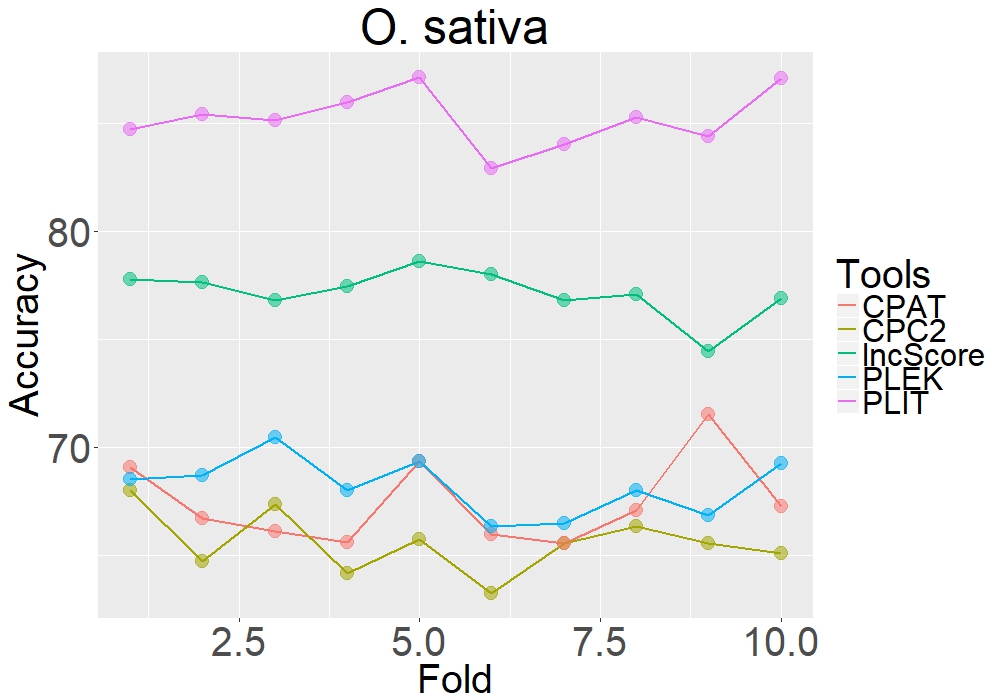}
  \centerline{(d)}
  \label{fig:sfig1}
\end{subfigure}
\begin{subfigure}{.5\textwidth}
  \centering
  \includegraphics[width=1.0\linewidth]{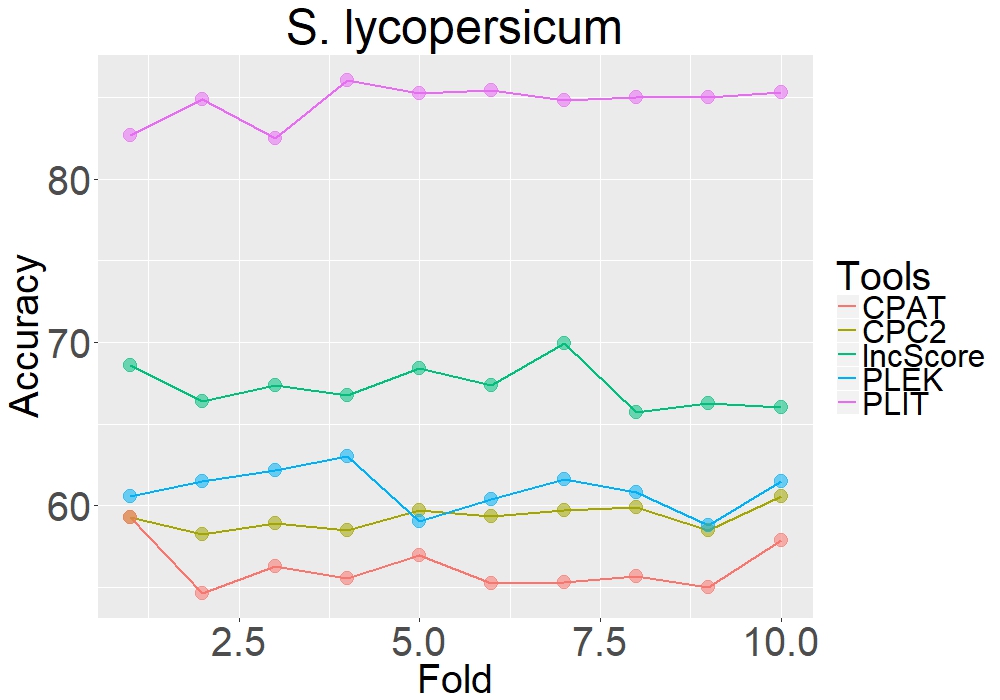}
  \centerline{(e)}
  \label{fig:sfig2}
\end{subfigure}
\begin{subfigure}{.5\textwidth}
  \centering
  \includegraphics[width=1.0\linewidth]{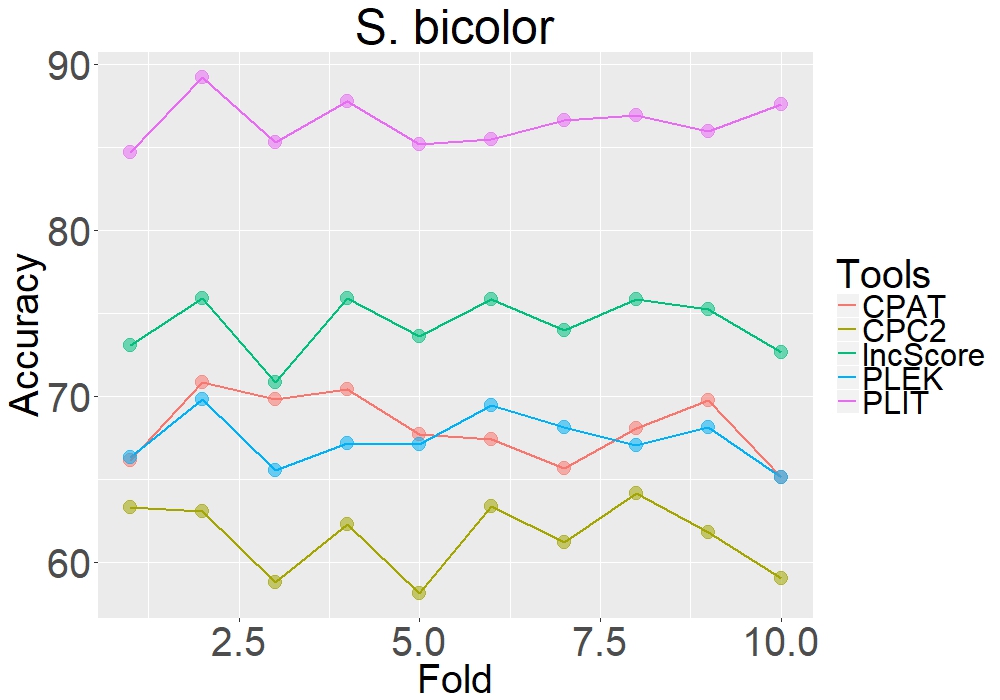}
  \centerline{(f)}
  \label{fig:sfig1}
\end{subfigure}
\begin{subfigure}{.5\textwidth}
  \centering
  \includegraphics[width=1.0\linewidth]{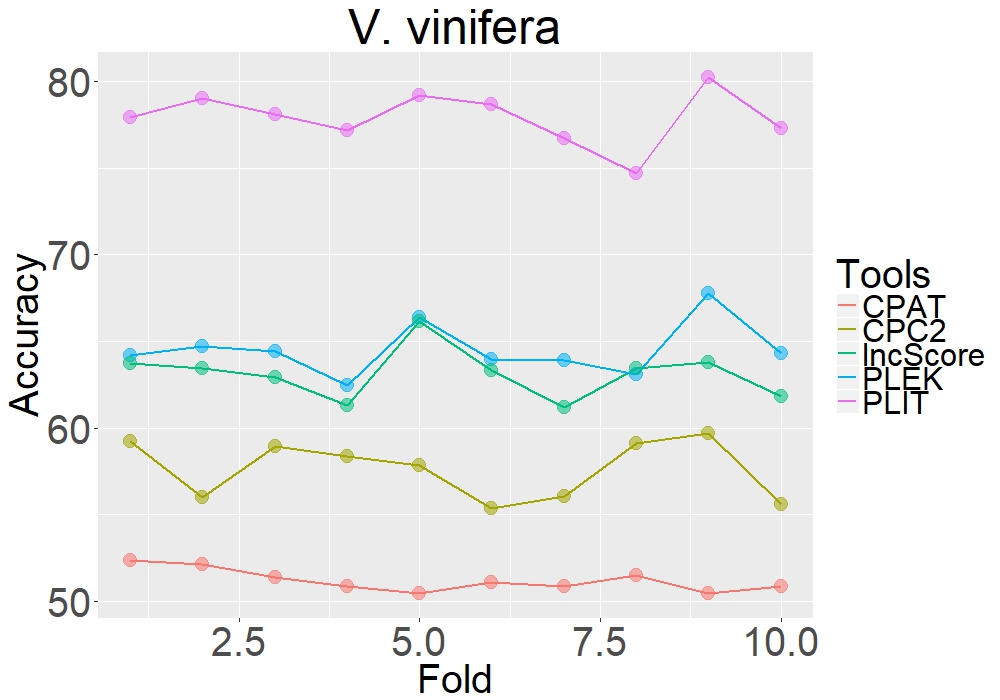}
  \centerline{(g)}
  \label{fig:sfig2}
\end{subfigure}
\begin{subfigure}{.5\textwidth}
  \centering
  \includegraphics[width=1.0\linewidth]{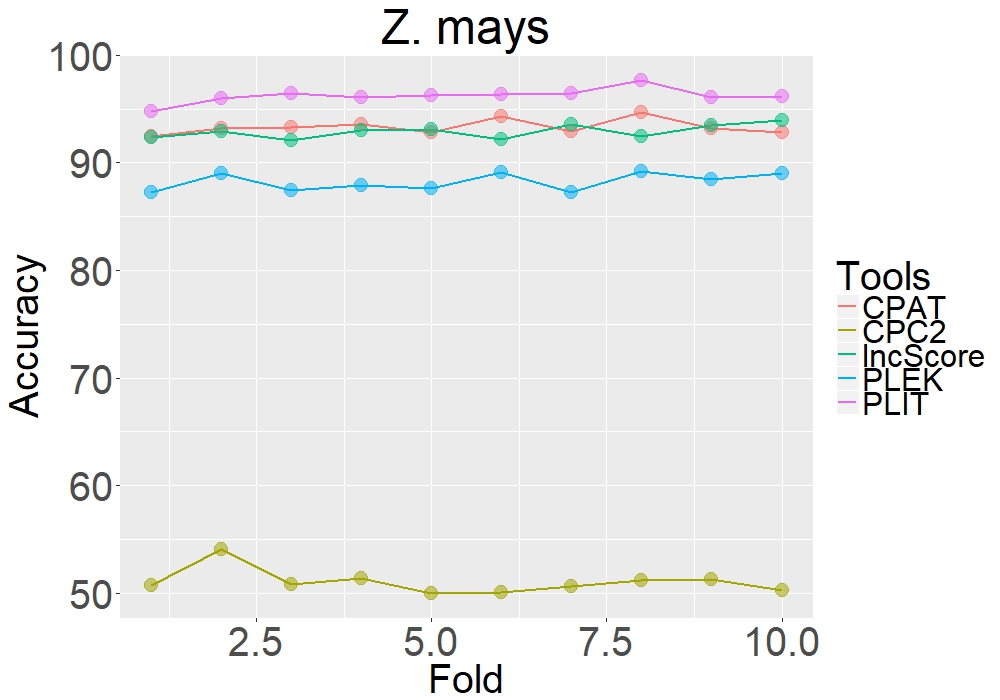}
  \centerline{(h)}
  \label{fig:sfig2}
\end{subfigure}%
\caption{Plots illustrating performance of PLIT against other existing tools based on 10-fold Cross Validation benchmarking analysis for (a) \textit{A. thaliana}, (b) \textit{A. thaliana}-EST derived lncRNA sequences, (c) \textit{G. max}, (d) \textit{O. sativa}, (e) \textit{S. lycopersicum}, (f) \textit{S. bicolor}, (g) \textit{V. vinifera}, and (h) \textit{Z. mays}. X-axis represents folds whereas y-axis represents percentage accuracy.}
\label{fig:fig}
\end{figure}

\subsection{Results of repeated 10-Fold Cross Validation performance benchmarking with data shuffling}

To further evaluate the efficiency and robustness of PLIT tool, a repeated 10-fold CV benchmarking was performed by repeatedly shuffling the coding and long non-coding sequences in each iteration. The prediction accuracies in each iteration were averaged to calculate a mean accuracy value along with standard error around the mean value (Figure~7\vphantom{\ref{fig:07}}). PLIT generated a mean accuracy 78.07\% with SE of 1.2, whereas lncScore, CPC2, PLEK and CPAT displayed mean accuracies of 68.41\%, 50.7\%, 53.5\% and 63.15\% along with SE ranging between 0.7 to 1.3 in \textit{A. thaliana} (Figure~7a and b\vphantom{\ref{fig:07}}). PLIT showed mean prediction accuracy range of 84.5 - 86.6\% for \textit{G. max}, \textit{O. sativa}, \textit{S. lycopersicum} and \textit{S. bicolor} datasets (Figure~7c, d, e and f\vphantom{\ref{fig:07}}), whereas for \textit{V. vinifera} and \textit{Z. mays} (Figure~7g and h\vphantom{\ref{fig:07}}), mean values of 78\%  and 96.2\% were obtained respectively. The SE value ranged between 1.2 and 1.7 for all the plant species except \textit{Z. mays} which displayed a SE of 0.45.

Prediction accuracies of CPC2, CPAT and lncScore over different test set sequences in several plant species displayed an average SE of 1.8 for \textit{G. max}, \textit{O. sativa} and \textit{S. bicolor}. In \textit{S. lycopersicum} dataset, an average SE of 1.38 was observed. For \textit{V. vinifera}, CPC2, PLEK and lncScore generated SE of 1.5, whereas PLEK produced a lower SE value of 0.5. The mean accuracy and SE plots demonstrates the consistency of accuracy values of PLIT across several folds and repetitions when tested against currently known and popular tools. As mentioned previously in Section 3.3, the difference in the accuracy remained consistently similar with slight deviation along the mean value.

\begin{figure}
\begin{subfigure}{.5\textwidth}
  \centering
  \includegraphics[width=1.0\linewidth]{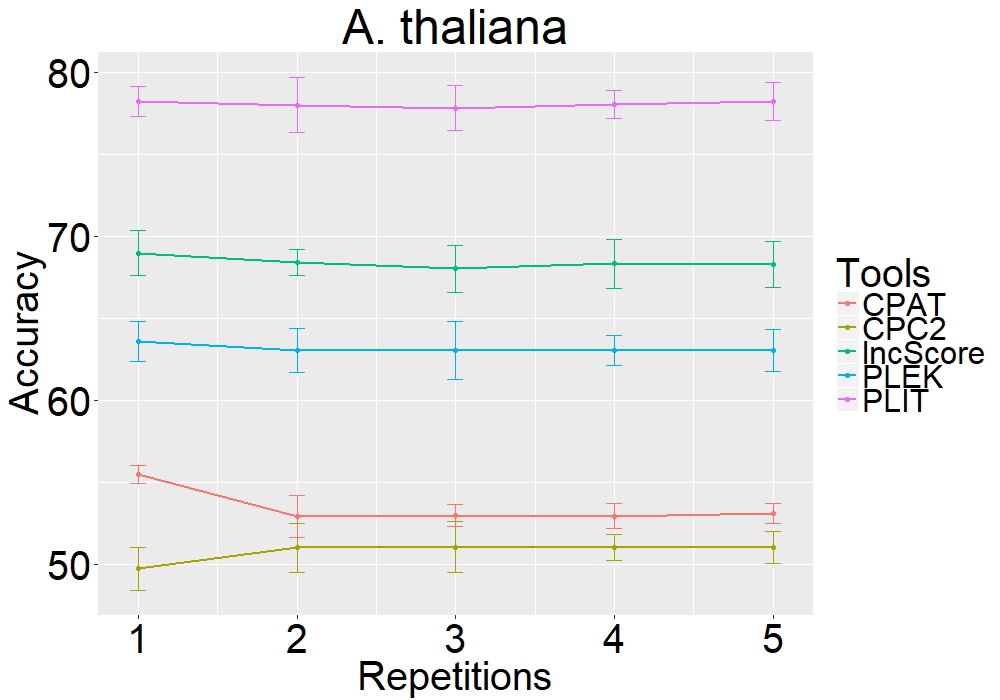}
  \centerline{(a)}
  \label{fig:sfig1}
\end{subfigure}%
\begin{subfigure}{.5\textwidth}
  \centering
  \includegraphics[width=1.0\linewidth]{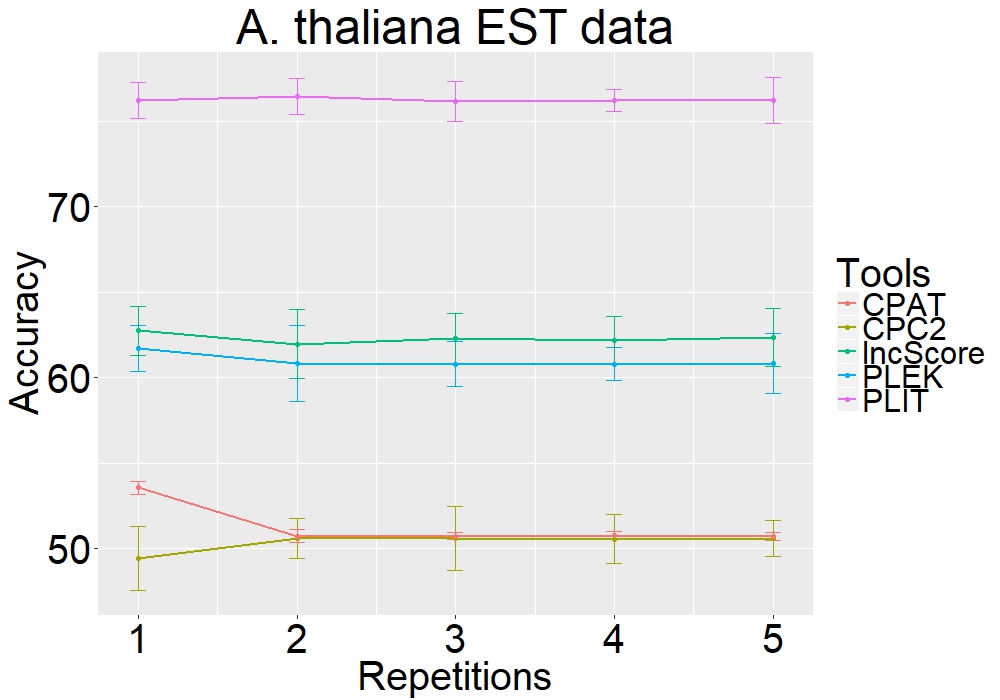}
  \centerline{(b)}
  \label{fig:sfig2}
\end{subfigure}
\begin{subfigure}{.5\textwidth}
  \centering
  \includegraphics[width=1.0\linewidth]{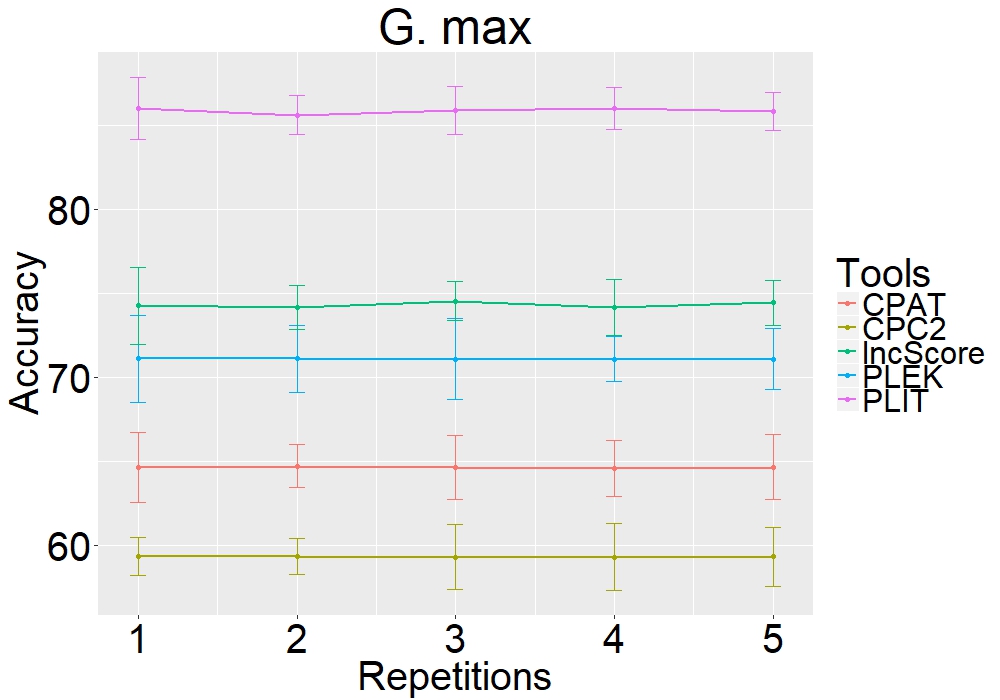}
  \centerline{(c)}
  \label{fig:sfig2}
\end{subfigure}
\begin{subfigure}{.5\textwidth}
  \centering
  \includegraphics[width=1.0\linewidth]{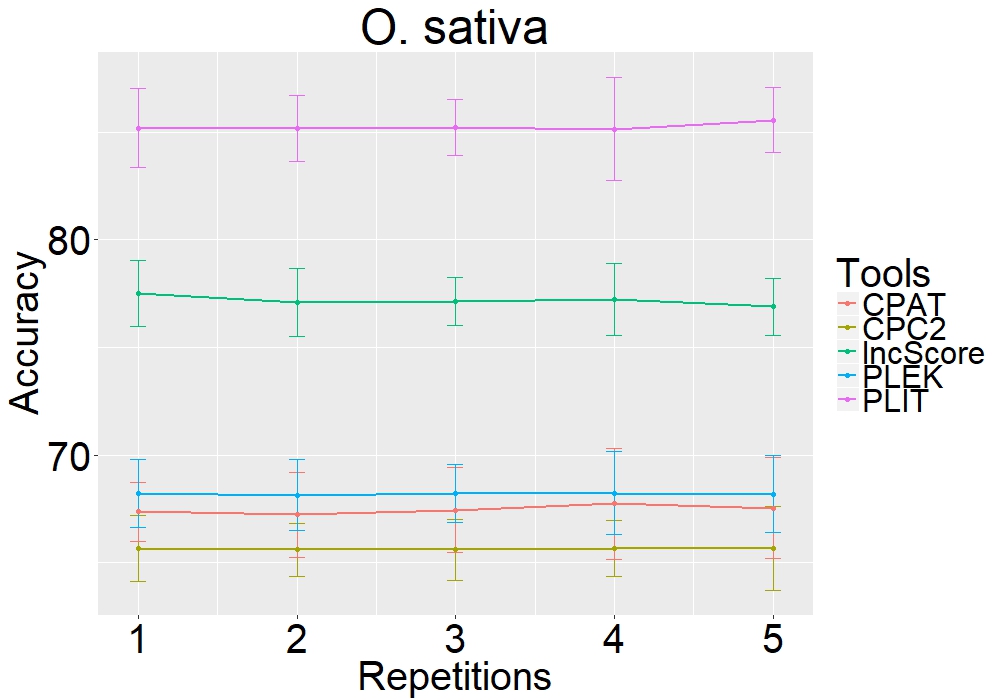}
  \centerline{(d)}
  \label{fig:sfig2}
\end{subfigure}
\begin{subfigure}{.5\textwidth}
  \centering
  \includegraphics[width=1.0\linewidth]{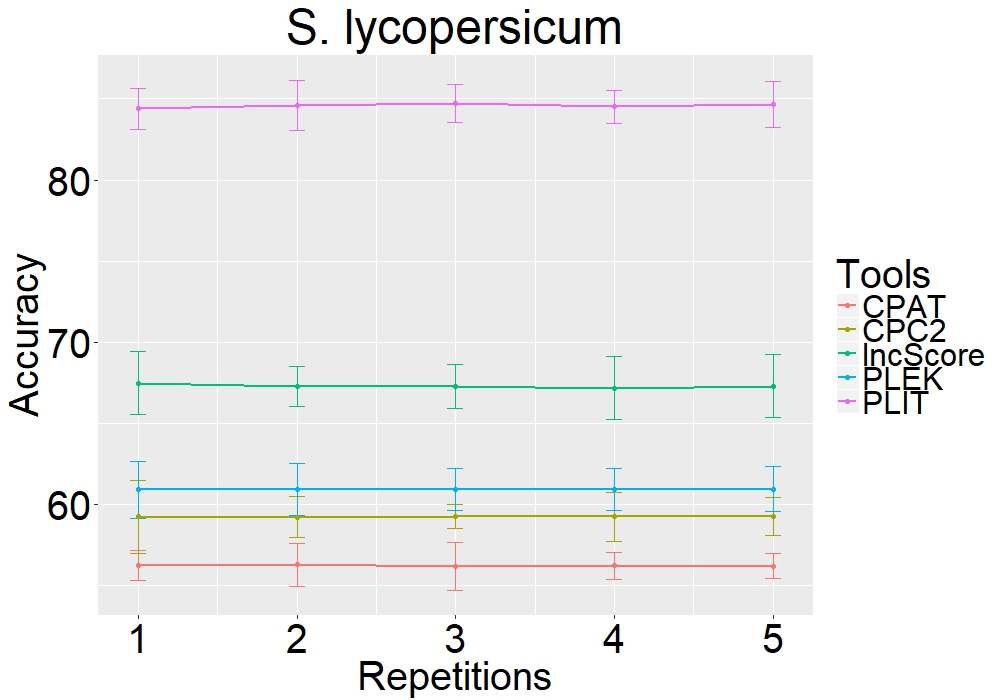}
  \centerline{(e)}
  \label{fig:sfig2}
\end{subfigure}
\begin{subfigure}{.5\textwidth}
  \centering
  \includegraphics[width=1.0\linewidth]{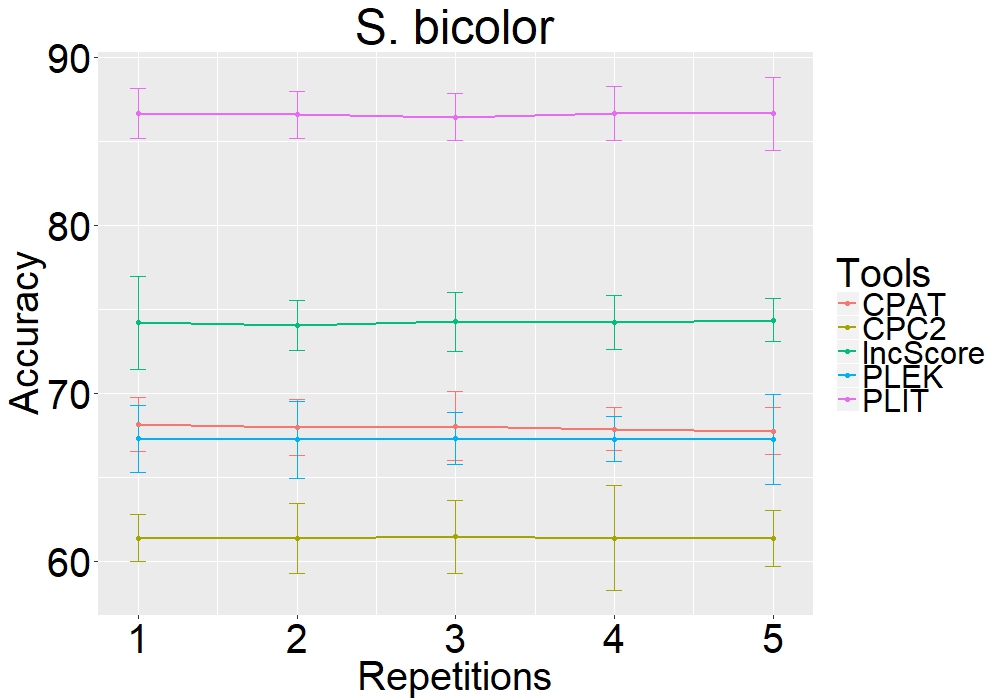}
  \centerline{(f)}
  \label{fig:sfig2}
\end{subfigure}
\begin{subfigure}{.5\textwidth}
  \centering
  \includegraphics[width=1.0\linewidth]{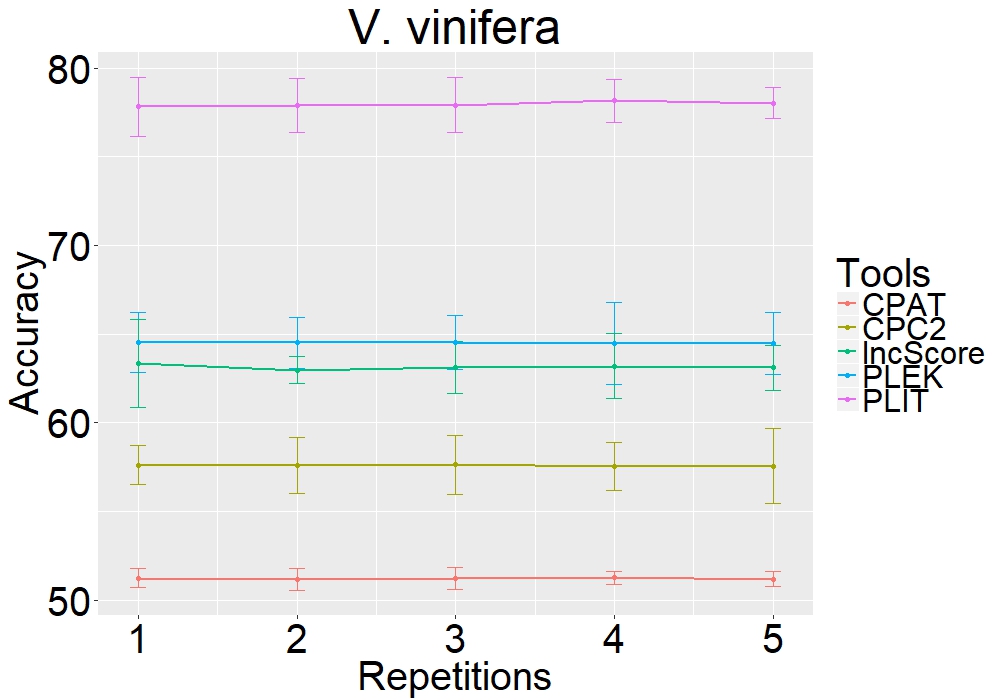}
  \centerline{(g)}
  \label{fig:sfig2}
\end{subfigure}
\begin{subfigure}{.5\textwidth}
  \centering
  \includegraphics[width=1.0\linewidth]{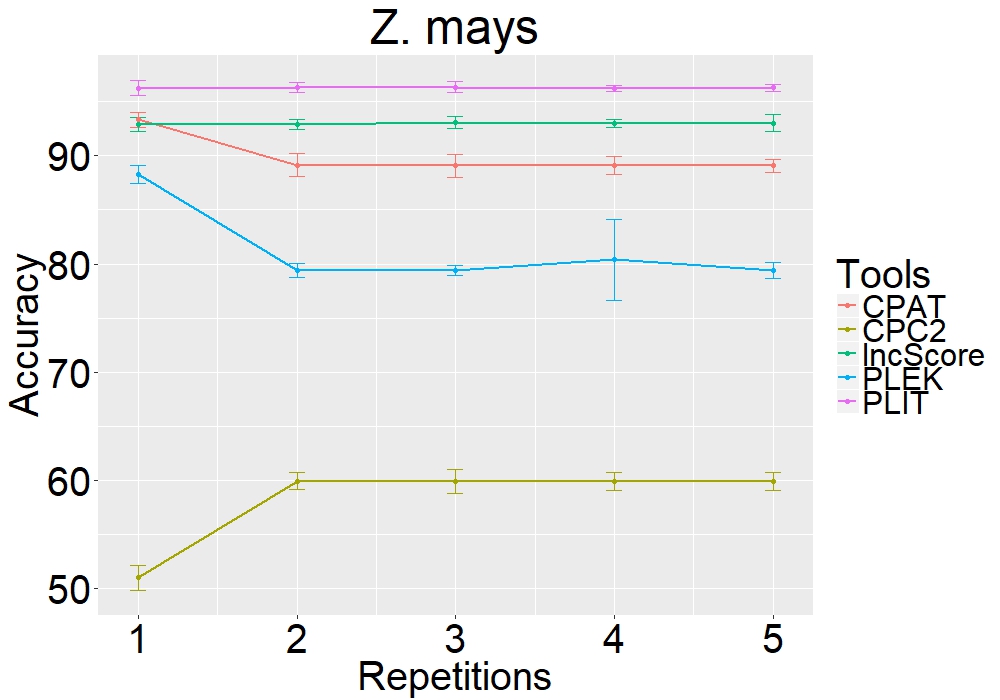}
  \centerline{(h)}
  \label{fig:sfig1}
\end{subfigure}%
\caption{Plots illustrating repeated 10-fold Cross Validation benchmarking analysis with data shuffling. Five repetitions have been performed on each RNA-seq dataset. Prediction accuracy of PLIT has been becnhmarked against CPAT, CPC2, PLEK and lncScore tools by averaging the accuracies over the folds and calculating the mean value on each repetition. The error bars indicate the Standard Error around the mean value. The following RNA-seq datasets have been used for repeated 10-fold CV analysis: (a) \textit{A. thaliana}, (b) \textit{A. thaliana}-EST derived lncRNA sequences, (c) \textit{G. max}, (d) \textit{O. sativa}, (e) \textit{S. lycopersicum}, (f) \textit{S. bicolor}, (g) \textit{V. vinifera}, and (h) \textit{Z. mays}. X-axis represents repetitions whereas y-axis represents percentage accuracy.}
\label{fig:fig}
\end{figure}

\subsection{Comparison of PLIT-LiRFFS against mRMR feature selection method}

The results of LiRFFS-based feature selection implemented in PLIT were compared against the feature set obtained from the mRMR (minimum-redundancy maximum relevancy) method. Using Mutual Information Difference (MID) in mRMR, 31 feature set was obtained on the unified dataset of 6 plant species. mRMR selected transcript length among ORF and sequence-based features. The remaining 30 RSCU features were selected by mRMR-based feature selection. Whereas LiRFFS selected 7 ORF and sequence-based features and 5 codon-biased and 19 RSCU-based codon-biased features were selected. The results of the feature selection from LiRFFS and mRMR approaches were applied and tested on five species: \textit{G. max}, \textit{O. sativa}, \textit{S. lycopersicum}, \textit{S. bicolor}, and \textit{v. vinifera}. Tables 6, 7 and 8 demonstrate comparison of prediction accuracy, sensitivity and specificity on different plant RNA-Seq datasets. The results clearly demonstrate a significant increase in prediction accuracy of 6.19\% in \textit{O. sativa}, 6.4\% in \textit{S. lycopersicum} and 4.7\% in \textit{S. bicolor} datasets, whereas a slightly higher increase of 0.36\% and 1.74\% for \textit{G. max} and \textit{V. vinifera} datasets. Sensitivity and specificity analysis results exhibits higher metric values in all the species except \textit{G. max} and \textit{V. vinifera} where a minor increase of 1.06\% in Sensitivity and 0.25\% was observed. 

\begin{table}[h!]
\centering
\caption{Performance comparison (percentage accuracy) of LiRFFS algorithm against mRMR-based selected features implemented on different plant RNA-seq datasets.}
\begin{tabular}{l@{\hspace{0.3cm}}l@{\hspace{0.6cm}}l@{\hspace{0.4cm}}l}
  \hline
Data set & LiRFFS Accuracy & mRMR Accuracy\\
  \hline
\textit{Glycine max} & 83.27 & 82.91\\
\textit{Oryza sativa} & 86.91 & 80.72\\ 
\textit{Solanum lycopersicum} & 84.30 & 77.90\\
\textit{Sorghum bicolor} & 85.92 & 81.22\\
\textit{Vitis vinifera} & 79.42 & 77.68\\
 \hline
\end{tabular}
\label{Tab:01}
\end{table}

\begin{table}[h!]
\vspace{-0.2cm}
\centering
\caption{Sensitivity comparison of LiRFFS algorithm against mRMR-based selected features implemented on different plant RNA-seq datasets.}
\begin{tabular}{l@{\hspace{0.3cm}}l@{\hspace{0.6cm}}l@{\hspace{0.4cm}}l}
  \hline
Data set & LiRFFS Sensitivity & mRMR Sensitivity\\
  \hline
\textit{Glycine max} & 81.85 & 82.91\\
\textit{Oryza sativa} & 86.91 & 82.54\\ 
\textit{Solanum lycopersicum} & 83.25 & 76.74\\
\textit{Sorghum bicolor} & 81.22 & 81.63\\
\textit{Vitis vinifera} & 75.06 & 71.32\\
 \hline
\end{tabular}
\label{Tab:01}
\end{table}

\begin{table}[h!]
\vspace{-0.2cm}
\centering
\caption{Specificity comparison of LiRFFS algorithm against mRMR-based selected features implemented on different plant RNA-seq datasets.}
\begin{tabular}{l@{\hspace{0.3cm}}l@{\hspace{0.6cm}}l@{\hspace{0.4cm}}l}
  \hline
Data set & LiRFFS Specificity & mRMR Specificity\\
  \hline
\textit{Glycine max} & 84.69 & 82.91\\
\textit{Oryza sativa} & 86.91 & 78.91\\ 
\textit{Solanum lycopersicum} & 85.34 & 79.07\\
\textit{Sorghum bicolor} & 90.61 & 80.81\\
\textit{Vitis vinifera} & 83.79 & 84.04\\
 \hline
\end{tabular}
\label{Tab:01}
\end{table}

\section{Discussion}

As RNA-Seq technology has been widely developed for identification of novel lncRNAs, recent advances in computer science has enabled computational prediction of lncRNAs solely from genomic sequence. Many tools developed for identification of lncRNAs paid less attention for their identification of the sequences derived from RNA-sequencing studies in plants. Therefore, it is necessary to develop a tool which could identify these long non-coding sequences with higher accuracy. In this study, we developed a new tool, called PLIT for accurately identifying the lncRNA transcript sequences from RNA-seq datasets. PLIT constructs a set of sequence-based and codon-bias features from the target FASTA sequence derived from RNA-seq data. The tool additionally implements and provides a feature selection-based approach for identifying a set of optimal features using LASSO and iRF classifier methods. The implementation of the LiRFFS algorithm in PLIT selected 24 codon-bias and 7 sequence features. The selected features shows that identification of lncRNA sequences primarily depends on open reading-frames, length of transcript, coverage of the ORFs in different sequences, frequency of hexamers, base pair positioning in the reading frames, GC content, contribution of frequency of base pairs in the codons and selection of synonymous codons in the transcript sequences. Furthermore, some of the extracted features such as GC content, sequence length as well as the ORF length which are implemented in PLIT have been previously administered in PredCircRNA tool for classification of circular RNAs from other lncRNA sequences \cite{Pan2015}. Another study conducted by Hu et al. (2015) developed RNAfeature for characterizing novel ncRNAs by finding significant features shared by various ncRNA sequences \cite{Hu2015}. The study determined 10 essential features including structures, sequences, expression profiles and histone modification signals. When compared with RNAfeature, the common feature implemented in both PLIT, PredCircRNA and RNAfeature tools is the GC content. This feature signifies its importance in lncRNA identification. Apart from GC content, RNAfeature used DNA and protein sequence conservation features, RNA secondary structure stability, homologs, conservation features and ORF property feature which is constructed from multiple sequence alignment method.         

In this study, a set of distinct 8 plant species from the Refseq database have been used for individually testing the prediction performance of the feature set implemented in PLIT tool. Furthermore, a unified set of 6 plant species from the Refseq database was constructed containing model and non-model organisms which were used for obtaining optimal set of features. A trimmed (minimal) set of features from the feature selection results can be further used for obtaining accurately identifying lncRNA sequences in plant RNA-Seq datasets. Therefore, current study focuses on the accurate prediction and benchmarking against other widely known tools. lncRNA sequences deposited in CANTATAdb database have been used as negative samples whereas protein-coding genes deposited in Ensembl database have been used as positive samples for constructing training and test set sequences for benchmarking the prediction accuracy of PLIT tool. The optimal feature set is obtained by iteratively constructing a training model from the training sequence set and testing the model on the validation set of coding and non-coding sequences. Since, Refseq and Ensembl databases contains lncRNA and protein-coding sequences for model and non-model species, a supervised classification approach can be implemented on the reference set of sequences which can potentially generate a set of optimal features. This reduced set of features can be used as input in further downstream analysis in wide variety of RNA-Seq plant species for the accurate identification of lncRNA sequences from protein-coding genes.

Identification of lncRNA sequences primarily exhibits their selection based on the ORFs on the DNA strand and position of purines and pyramidines in the reading frame. Additionally, the effect of increased GC content in lncRNA prediction is linked to the codon usage pattern where higher GC content indicates heterogeneity \cite{Zhou2014}. A study conducted by Zhou et al. (2014) confirms the existence of linear relationship between amino acid usage and genomic GC content \cite{Zhou2014}. Another study conducted by Biro et al. (2008) discusses about the correlations between individual codons as well as codon residues at different codon positions \cite{Biro2008}. Furthermore, non-randomness of synonymous codon usage is highly affected by tRNA pool size having a primary role in the reading frame and imposing constraints on the synonymous codon usage \cite{Antezana1999,Lipman1983}.  

These results provide insights into the preferential selection of synonymous codons in the classification process. LiRFFS produced a minimal and maximal set of optimal feature sets from the training and validation datasets constructed from six plant species. The AUC profiles of the 31F optimal feature set on plant RNA-seq datasets demonstrates comparatively higher performance when compared against the 7F set. The similarity of the ROC curves for 31F and 73F sets indicate better selection of features represented by greater prediction performance. Test set sequences used in RNA-Seq datasets were used to demonstrate prediction accuracy of PLIT tool against other existing tools and its application for identifying novel lncRNAs based on the optimal feature set. 

When comparing against currently popular state of the art alignment-free tools such as CPAT, CPC2, PLEK and lncScore \cite{Kang2017,Wang13,Li14,JZhao16}, PLIT generated much better prediction accuracy values when tested on several plant datasets. From 10-fold CV and repeated 10-fold CV analyses, it was found that the prediction accuracies of other tools were comparitively lower with average differences ranging between 9 to 30\%. The sensitivity and specificity values obtained from other tools displayed a biased prediction thereby generating greater false negatives or false positives. Prediction performance of PLIT showed accuracies $>$80\% on most plant species whereas other tools performed inconsistently with accuracies varying significantly across different species.

Compared with other tools, PLIT provides unique set of features for accurate identification of lncRNAs transcripts which provides several advantages over other tools. First, apart from commonly known distinguishing sequence-based features such as ORF length, ORF coverage, GC content, Fickett score and Hexamer score, it takes advantage of codon-biased features to increase its discriminative power. Second, PLIT implements a powerful semi-supervised optimization approach for selection of principal features which can be applied on plants and vertebrates. Third, the implemented LiRFFS and random forest classifier for feature selection and prediction of lncRNAs offers robustness, efficiency and suitability on multiple model as well as non-model plant species. The results of PLIT revealed higher accuracy with balanced sensitivity and specificity on all test sets. This implies that the models generated by PLIT did not overfit the training data.

\section{Conclusions}

In this work, we developed a novel tool, PLIT, for accurate identification and discovery of lncRNA sequences particularly well-suited for RNA-seq data from plants. PLIT exceeds the prediction performance over other tools on various parameters. The ability to identify and differentiate various lncRNA transcripts was demonstrated with several cross-validation tests on different RNA-seq datasets. Using LiRFFS, optimal features were identified based on the FASTA sequences from Refseq database. Evaluation with K-fold and repeated K-fold CV demonstrated consistent and superior performance of PLIT on all plant datasets against other tools. Thus, PLIT is a stable, robust and accurate tool for distinguishing long non-coding and protein-coding transcripts from plant RNA-seq data.    

\section*{Acknowledgements}

This work has been sponsored by Coventry University; Faculty of Engineering, Environment, and Computing; School of Computing, Electronics, and Mathematics.

\section*{References}

\bibliography{mybibfilePLIT}

\end{document}